\documentclass[aps,pra,twocolumn,showpacs,superscriptaddress,amsart]{revtex4}
\pdfoutput=1
\usepackage{graphics}
\usepackage{amssymb}
\usepackage{amsmath}
\usepackage{epsfig}
\usepackage{color}
\usepackage{verbatim}
\usepackage{float}
\usepackage{mathrsfs}
\usepackage{esint}
\usepackage{hyperref}
\usepackage{tabularx}
\usepackage{multirow}
\begin{document}

\title{Superfluid phases of fermions with hybridized $s$ and $p$ orbitals}
\author{Shaoyu Yin}
\author{J. E. Baarsma}
\author{M. O. J. Heikkinen}
\author{J.-P. Martikainen}
\address{COMP Centre of Excellence, Department of Applied Physics, Aalto University, FI-00076 Aalto, Finland}
\author{P. T\"orm\"a}\thanks{paivi.torma@aalto.fi}
\address{COMP Centre of Excellence, Department of Applied Physics, Aalto University, FI-00076 Aalto, Finland}
\address{Institute for Quantum Electronics, ETH Z\"urich, 8093 Z\"urich, Switzerland}

\begin{abstract}
We explore the superfluid phases of a two-component Fermi mixture with hybridized orbitals in optical lattices. We show that there exists a general mapping of this system to the Lieb lattice. By using simple multiband models with hopping between $s$ and $p$-orbital states, we show that superfluid order parameters can have a $\pi$-phase difference between lattice sites, which is distinct from the case with hopping between $s$-orbitals. If the population imbalance between the two spin species is tuned, the superfluid phase may evolve through various phases due to the interplay between hopping, interactions and imbalance. We show that the rich behavior is observable in experimentally realizable systems.
\end{abstract}
\pacs{03.75.Ss, 05.30.Fk, 74.20.Fg, 71.10.Pm}
\maketitle

\section{Introduction}
\label{sec:introduction}
Multiband effects are important in understanding a variety of quantum many-body phenomena such as high-temperature superconductivity,
fractional quantum Hall phases, and topological matter in general \cite{nature_multibandsc_2001,Hasan_topological_review2010,kopnin_high-temperature_2011,Bernevig_topologicalbook2013}. In the context of ultracold gases, excellent opportunities of experimentally exploring multiband phenomena in a
controlled way have emerged in the recent years. Optical lattices naturally have multiple bands, and quantum systems in the excited bands of optical lattices have
been experimentally realized. M\"{u}ller {\it et al.}
~\cite{muller_state_2007} transferred ultracold bosons
into the $p$-band of the lattice and observed how coherence
was established between atoms. Recently Zhai {\it et al.} prepared bosons in the
$d$-band~\cite{zhai_effective_2013}.
Closely related to the work presented in this article are the experiments~\cite{wirth_evidence_2011,
olschlager_unconventional_2011,olschlager_interaction-induced_2013}
exploring excited band condensates in a bipartite optical lattice, and \cite{Liu_piphase1D}, studying pairing between different parity orbital fermions. Bosons in the flat band of a Lieb lattice were realized very recently \cite{taie_matter-wave_2015}.

These experimental possibilities have insprired considerable amount of theoretical work on multiband effects in optical lattices. Naturally, in other contexts 
the volume of work on multiband effects is much larger; here we mention only examples of results related to ultracold gases. 
For example, it has been argued that since higher bands of an optical lattice
typically have larger bandwidths, higher critical temperatures for anti-ferromagnetic ordering may be realized \cite{wu_theory_2008}. The idea is probably mentioned in many places and is based on the fact that in perturbation theory
the Hamiltonian ends up having a prefactor tunneling squared over coupling. 
In the $p$-band tunneling can be larger so the chracteristic energy over temperature scale can be higher in absolute sense.
(Quote from Wu {\it et al.} \cite{wu_theory_2008}: ``We also show that in
the strongly correlated regime the N\'{e}el temperature for $p$ band antiferromagnetism is 2 to 3 orders of magnitudes
higher than that of $s$ band, which is much more promising to be attained in cold atom experiments.")
It has also been suggested by Dutta {\it et al.}~\cite{dutta_emergent_2014} that
strong interactions and multiband effects can give rise to
self-assembly of non-trivial lattices for topological insulators.
Topological semimetals and chiral
superfluidity with $s$-wave interactions have been predicted~\cite{sun_topological_2012,liu_chiral_2014}
in multi-orbital models where orbitals with different symmetries
interact. There is also extensive theory literature on both
$p$-band bosons~\cite{isacsson_multiflavor_2005,liu_atomic_2006,collin_quantum_2010,li_time-reversal_2012}
and fermions~\cite{wu_pxy-orbital_2008} with many studies focusing on the strong coupling
regime~\cite{zhao_orbital_2008,pinheiro_xyz_2013}.
Furthermore, oscillating  order parameters
in fermionic systems have been predicted due to
coupling between $s$- and $p$ orbitals~\cite{zhang_modulated_2010},
or in pure $p$-orbital systems~\cite{WuDasSarma_fwave2010,cai_stable_2011}.

Motivated by these advances, in this article we explore the physics of attractively
interacting two-component fermions with multiple bands.
In particular, we wish to understand how an unequal number of
different fermionic species and the tunneling properties of $p$-orbitals
influence the formation of $s$-wave pairing order parameters in such systems.

We first solve a simplified model with just two sites which indicates
possibilities of different superfluid phases, some with a spatially varying order parameter phase factor.
We then demonstrate that in many respects, the
results from this toy model are realized by a bipartite lattice where
$s$ and $p$-orbitals in different sublattices hybridize~\cite{martikainen_multiorbital_2012}.
Such a lattice has been experimentally demonstrated
by Wirth {\it et al.}~\cite{wirth_evidence_2011} who
studied Bose-Einstein condensation on the excited bands
of such a lattice and found non-trivial ordering of the condensate
phase. This ordering is due to the fairly complex interplay between
tunneling properties of different orbitals and on-site interactions
between atoms.

We outline the expected phase diagram for fermions
at the mean-field level and find a possibility of a $\pi$-phase superfluidity where the sign
of the order parameter varies between sublattices. Such possibility
was raised by Iskin~\cite{iskin_superfluid_2013}
in the context of a checkerboard lattice, but it turned out that
in that system this possibility was not realized. Somewhat related phenomena have also been discussed in studies
exploring FFLO phases in lattices~\cite{koponen_finite-temperature_2007,loh_detecting_2010,kim_exotic_2011,chiesa_phases_2013},
in multiorbital models~\cite{zhang_modulated_2010,cai_stable_2011}, or in two-dimensional systems
without a lattice~\cite{yin_fulde-ferrell_2014}.

This paper is organized as follows: We start, in Sec.~\ref{sec:system},
by discussing the lattice we consider in some detail and we especially elaborate on the sign changes that occur in the hopping parameters, due to the different orbital states. Consequently, in Sec.~\ref{sec:pairing}, we introduce interactions and the possibility of a pairing instability. We first study a simplified, dispersionless model to examine what kind of pairing instabilities can occur.
Subsequently, in Sec. IV, we study a more complex system, resembling an experimentally realizable system. The numerical results are presented in Sec. V. Finally in Sec. VI, we conclude with a summary and discussion.

\section{Simple model with hybridized orbitals}
\label{sec:system}
We consider fermions in two different (pseudo) spin states, $\uparrow$ and $\downarrow$, occupying bipartite lattices with sublattices $A$ and $B$. Here, the fermions on sites of the $A$ sublattice are in the $s$-orbital state, whereas the particles in the $B$ sublattice occupy a $p$-wave orbital state, where the number of the different $p$-orbitals is equal to the dimension of the lattice. In this section, we study first non-interacting fermions occupying a one-dimensional and consequently a two-dimensional lattice, where we denote the orbitals on the $B$ sublattice by $p$ or $p_x$ and $p_y$, respectively. Our goal here is to elucidate how systems with hybridized $s$ and $p$  orbitals are connected to and differ from systems with $s$ orbitals only.

\subsection{One dimension}
\begin{figure}
\includegraphics[width=.9\columnwidth]{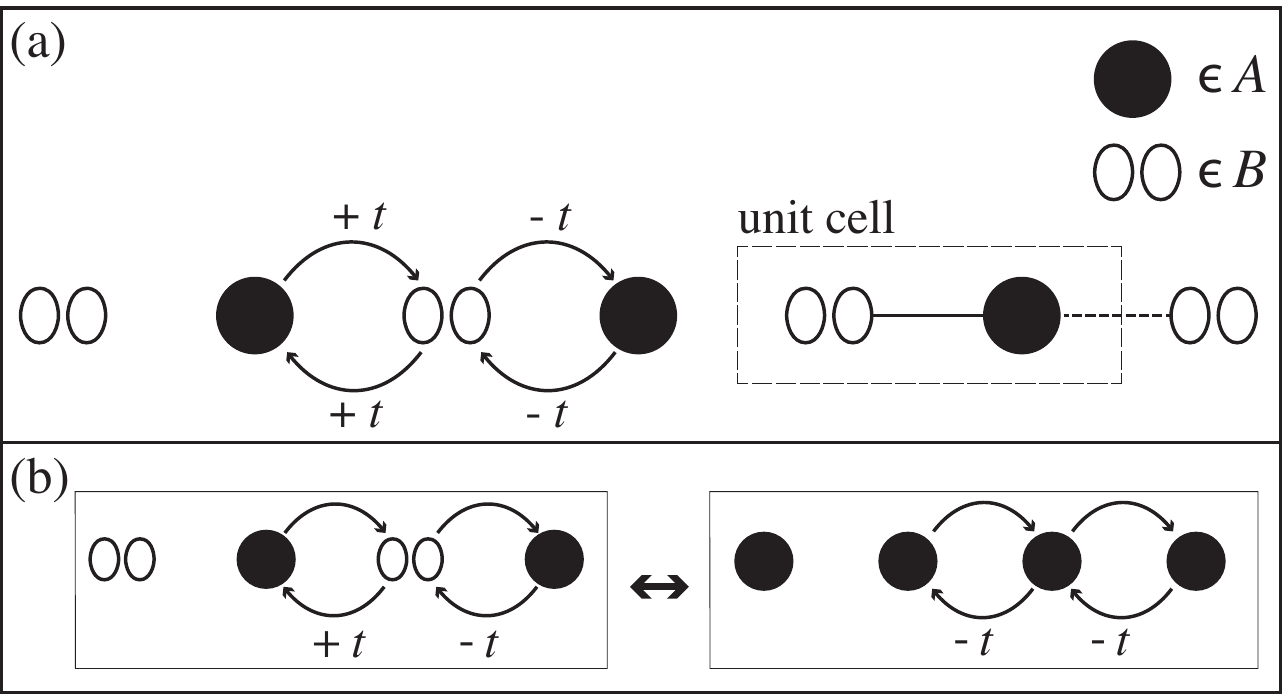}
\caption{The lattice we study in one dimension (a), where on the $B$ lattice sites the $p$ orbitals are sketched. Hopping within the specified unit cell is denoted by a full line, whereas hopping to the neighboring unit cell is denoted by a dashed line. In (b) the mapping from the $s\text -p$ lattice to the lattice with $s$ orbitals only is shown.}
\label{1Dlattices}
\end{figure}

Due to the different orbital states on the two sublattices, the hopping coefficients are also different for particles moving in opposite directions. To be more specific, we first focus on the one-dimensional case. There, the hopping coefficient for a particle moving from an $A$ to a neighbouring $B$ site in the positive $x$-direction, $t^{sp}_{+x}$, has an opposite sign to the one in the opposite direction, $t^{sp}_{-x}$, which is due to the odd parity of the $p$ orbital. The hopping coefficients for the particles moving back from the $B$ to the $A$ sublattice are the same, $t^{sp}_{+x}=t^{ps}_{-x}$ and $t^{sp}_{-x}=t^{ps}_{+x}$, since they correspond to the same overlap integral. Apart from the sign difference the hopping coefficients are the same, $t^{sp}_{+x}=-t^{sp}_{-x}\equiv t$.
Thus, the nearest neighbour hopping Hamiltonian for the one-dimensional case reads
\begin{align}
H_K^{1D}=-t\sum_{n,\sigma}\left[\hat{\psi}^{s\dagger}_{\sigma,n}\hat{\psi}^p_{\sigma,n}-\hat{\psi}^{p\dagger}_{\sigma,n+1}\hat{\psi}^s_{\sigma,n}+\text{h.c.}\right],
\label{1dspHamiltonianx}
\end{align}
where $\hat{\psi}_{\sigma,n}^{j\dagger}$ creates a (pseudo) spin state $|\sigma\rangle$ fermion with orbital $j$ at unit cell $n$ and a unit cell contains one $A$ and one $B$ site, see Fig.\ref{1Dlattices}(a). The first term thus describes hopping within a unit cell, while the second describes hopping across unit cells.

It is instructive to Fourier transform the above Hamiltonian Eq.(\ref{1dspHamiltonianx}), in a few steps
\begin{align}
\nonumber H_K^{1D} =-t\sum_{k,\sigma} &\left[\hat{\psi}^{s\dagger}_{\sigma,k}\hat{\psi}^p_{\sigma,k}-e^{\text{i}k2d}\hat{\psi}^{p\dagger}_{\sigma,k}\hat{\psi}^s_{\sigma,k}+\text{h.c.}\right]\\
\nonumber =-t\sum_{k,\sigma} &\left[\left(1-e^{-\text{i}k2d}\right)\hat{\psi}^{s\dagger}_{\sigma,k}\hat{\psi}^p_{\sigma,k}\right.\\
\nonumber &+\left.\left(1-e^{\text{i}k2d}\right)\hat{\psi}^{p\dagger}_{\sigma,k}\hat{\psi}^s_{\sigma,k}\right]\\
 =-t\sum_{k,\sigma} & 2\text{i}\sin(kd)\left(\tilde{\psi}^{s\dagger}_{\sigma,k}\tilde{\psi}^p_{\sigma,k}
-\tilde{\psi}^{p\dagger}_{\sigma,k}\tilde{\psi}^s_{\sigma,k}\right),
\label{1dspHamiltoniank}
\end{align}
where $d$ is the lattice spacing, which is taken equal to one here. A transformation in the fermionic operators was made in the last line, $\tilde{\psi}^p_{\sigma,k}=\exp[-\text ikd]\hat{\psi}^p_{\sigma,k}$ and $\tilde{\psi}^s_{\sigma,k}=\hat{\psi}^s_{\sigma,k}$.

\subsection{Mapping $p$ to $s$ orbitals}
\label{sec:mapping}
Although the Hamiltonian in Eq.(\ref{1dspHamiltonianx}) and its Fourier transform in Eq.(\ref{1dspHamiltoniank}) look quite different from the Hamiltonian describing fermionic particles in a 1D lattice with only $s$-orbital sites, it turns out that these two are more similar than they seem. We show this connection by transforming the Hamiltonian with alternating hoppings to a hopping Hamiltonian without sign changes, see Fig.\ref{1Dlattices}(b). First, Eq.(\ref{1dspHamiltonianx}) can be rewritten by splitting the summation over the unit cells in a sum over the even and odd unit cells, after which the Hamiltonian reads
\begin{align}
\nonumber H_K^{1D}=-t\sum_{m,\sigma} & \left[\hat{\psi}^{s\dagger}_{2m,\sigma}\hat{\psi}^p_{2m,\sigma}+\hat{\psi}^{s\dagger}_{2m+1,\sigma}\hat{\psi}^p_{2m+1,\sigma}\right.\\
\nonumber &\left. -\hat{\psi}^{p\dagger}_{2m,\sigma}\hat{\psi}^s_{2m+1,\sigma}
-\hat{\psi}^{p\dagger}_{2m+1,\sigma}\hat{\psi}^s_{2m+2,\sigma}+\text{h.c.}\right],
\end{align}
where the summation over $m$ runs over half of the values that $n$ in Eq.(\ref{1dspHamiltonianx}) runs over. Consequently, the following unitary transformation can be used
\begin{align}
\tilde{\psi}^{j}_{2m,\sigma} &= \hat{\psi}^{j}_{2m,\sigma} \nonumber\\
\tilde{\psi}^{j}_{2m+1,\sigma} &= - \hat{\psi}^{j}_{2m+1,\sigma}, 
\label{evenoddtransformation}
\end{align}
with $j=s,p$. This yields for the Hamiltonian
\begin{align}
\nonumber H_K^{1D}=-t\sum_{m,\sigma} & \left[\tilde{\psi}^{s\dagger}_{2m,\sigma}\tilde{\psi}^p_{2m,\sigma}
+\tilde{\psi}^{s\dagger}_{2m+1,\sigma}\tilde{\psi}^p_{2m+1,\sigma}\right.\\
\nonumber &\left. +\tilde{\psi}^{p\dagger}_{2m,\sigma}\tilde{\psi}^s_{2m+1,\sigma}
+\tilde{\psi}^{p\dagger}_{2m+1,\sigma}\tilde{\psi}^s_{2m+2,\sigma}+\text{h.c.}\right]\\
\nonumber =-t\sum_{n,\sigma} & \left[\tilde{\psi}^{s\dagger}_{n,\sigma}\tilde{\psi}^p_{n,\sigma}
+\tilde{\psi}^{p\dagger}_{n,\sigma}\tilde{\psi}^s_{n+1,\sigma}+\text{h.c.}\right],
\end{align}
which is the hopping Hamiltonian for particles in a 1D lattice with only $s$-orbital sites. In Fourier space the same transformation boils down to making a shift in the quasi momentum of the operators, $\hat{\psi}^{j}_{k,\sigma}=\tilde{\psi}^{j}_{k-\pi/(2d),\sigma}$. After subsequently shifting all quasi momenta by $+\pi/(2d)$ the Hamiltonian reads
\begin{align}
\nonumber H_K^{1D}=-t\sum_{k,\sigma} &\left[\left(1-e^{-\text{i}k2d}e^{-\text i\pi}\right)\hat{\psi}^{s\dagger}_{k,\sigma}\hat{\psi}^p_{k,\sigma}\right.\\
\nonumber &+\left.\left(1-e^{\text{i}k2d}e^{\text i\pi}\right)\hat{\psi}^{p\dagger}_{k,\sigma}\hat{\psi}^s_{k,\sigma}\right]\\
=-t\sum_k & 2\cos(kd)\left(\tilde{\psi}^{s\dagger}_{k,\sigma}\tilde{\psi}^p_{k,\sigma}
+\tilde{\psi}^{p\dagger}_{k,\sigma}\tilde{\psi}^s_{k,\sigma}\right),
\label{1DssHamk}
\end{align}
where the same transformation as before has been used in the last step.

Although there thus exists a simple mapping between the $s\text -p$ Hamiltonian and the $s$ orbitals only lattice, from Eq.(\ref{1DssHamk}) it is also clear that they are not exactly the same, the difference being the dispersions of the particles. The different dispersions result in a number of differences between the two systems, such as different momentum distributions and Fermi momenta. The latter can in turn give rise to different properties of possible superfluid phases.

\subsection{Two dimensions}
\begin{figure}
\includegraphics[width=.9\columnwidth]{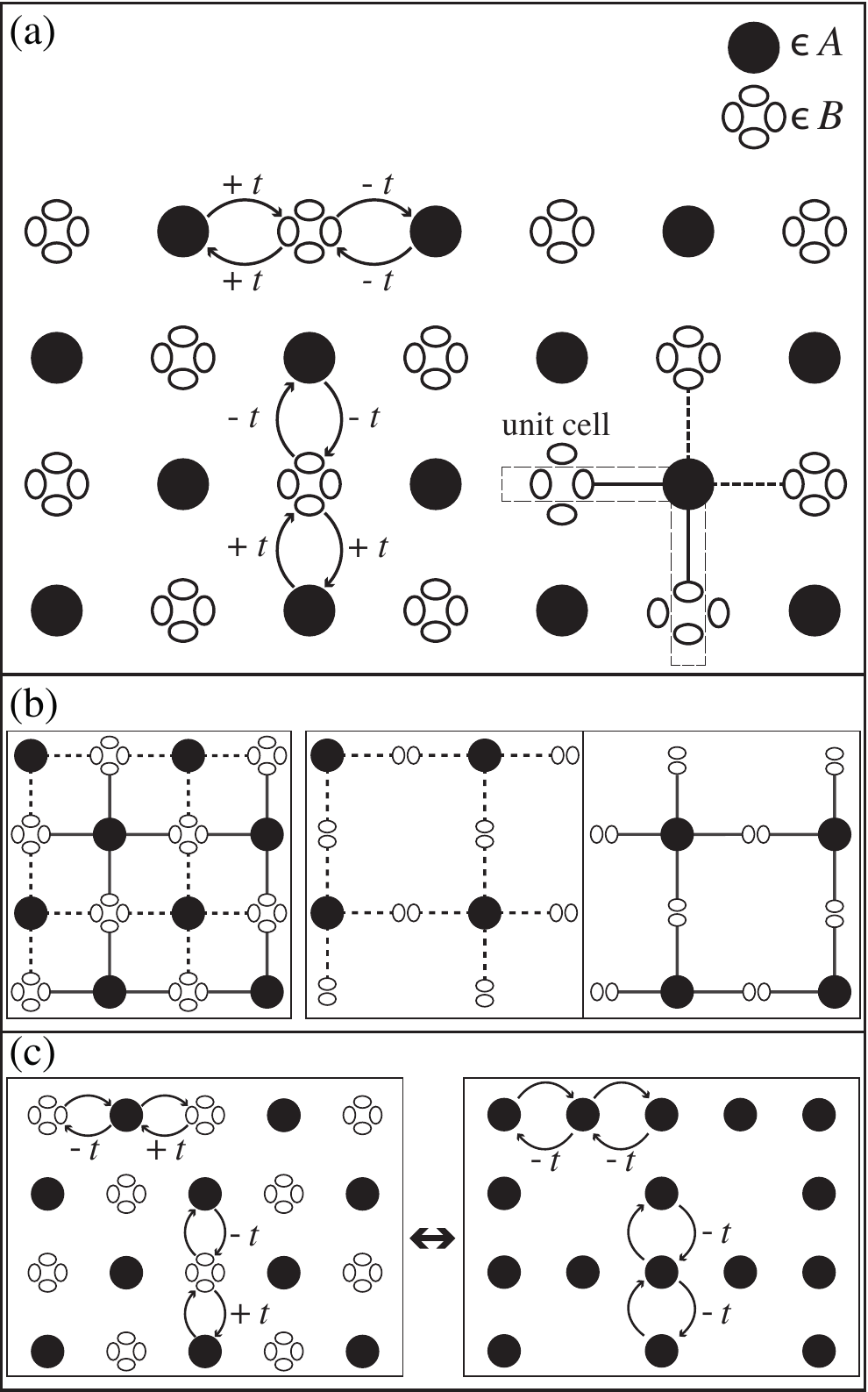}
\caption{(a) The $s\text -p$ lattice in two dimensions, where on the $B$ lattice sites the $p_x$ and $p_y$ orbitals are sketched. Hopping within the specified unit cell is denoted by a full line and to the neighboring unit cells by a dashed line. (b) Illustration of the two decoupled lattices. On the left the full lattice is sketched and the two different hopping routes for the particles are denoted by full and dashed lines, which are shown as separate lattices in the middle and right. (c) mapping from the $s\text -p$ lattice to the lattice with $s$ orbitals only.}
\label{simple2D}
\end{figure}

In a two-dimensional system the lattice sites of the $B$ sublattice contain two $p$-orbitals, $p_x$ and $p_y$ (denoted by $x$ and $y$ in sub- or superscripts, respectively), see Fig.\ref{simple2D}(a). The presence of two $p$-orbitals changes the hopping physics yet a bit more compared to the one-dimensional case. Namely, a particle that moves from the $A$ to the $B$ sublattice in the $x$ direction, ends up in a $p_x$ orbital state. Since the overlap integral between the $p_x$ orbital and the neighboring $s$-orbitals in the $y$-direction vanishes, this particle can only move along the $x$-direction. And similarly for particles moving from the $A$ to the $B$ sublattice in the $y$ direction. In other words, $t^{sy}_{\pm x}=t^{sx}_{\pm y}=0$ due to the odd parity of the $p$-orbitals. In the absence of interactions, this results in effectively two hopping sublattices, illustrated in Fig. \ref{simple2D}(b), which both have the Lieb lattice geometry. 
The hopping Hamiltonian in two dimensions reads
\begin{align}
H_K^{2D}=-t\sum_{n,\sigma;\alpha\in\{x,y\}}\left[\hat{\psi}^{s\dagger}_{\sigma,n}\hat{\psi}^{\alpha}_{\sigma,n}-\hat{\psi}^{\alpha\dagger}_{\sigma,n+\alpha}\hat{\psi}^s_{\sigma,n}+\text{h.c.}\right],
\label{2dspHamiltonianx}
\end{align}
where a unit cell contains one site from the $A$ and one from the $B$ sublattice, see Fig.\ref{simple2D}(a), and the second term in the summation now describes hopping to the next unit cell both in the $x$ and $y$ direction, denoted by $n+x$ and $n+y$ respectively.

The Fourier transform of the hopping Hamiltonian for the two-dimensional system is
\begin{align}
H_K^{2D}=-t\sum_{k,\sigma;\alpha\in\{x,y\}} & 2\text{i}\sin(k_\alpha d)\left(\tilde{\psi}^{s\dagger}_{\sigma,k}\tilde{\psi}^{\alpha}_{\sigma,k}
-\tilde{\psi}^{\alpha\dagger}_{\sigma,k}\tilde{\psi}^s_{\sigma,k}\right),
\label{2dspHamiltoniank}
\end{align}
where $d$ is again the lattice spacing and the same transformation in the fermionic operators has been used as in the one-dimensional case Eq.(\ref{1dspHamiltoniank}).

\subsection{Mapping $p_x$ and $p_y$ to $s$ orbitals}
Also in a two-dimensional system a mapping can be made from the $s\text-p$ lattice discussed above to a lattice containing only $s$-orbital states. The mapping from a square lattice containing both $s$ and $p$ orbitals is to the Lieb lattice containing only $s$ orbitals, see Fig.\ref{simple2D}(c). Actually, the two hopping sublattices both map to a Lieb lattice. Since they are uncoupled, the Hamiltonian in Eq.(\ref{2dspHamiltonianx}) can also be written as the sum of the two hopping Hamiltonians of the sublattices
\begin{align}
H_K^{2D}=H_{K1}+H_{K2},\nonumber
\label{2dspLieb}
\end{align}
where both $H_{K1}$ and $H_{K2}$ have the same form as Eq.(\ref{2dspHamiltonianx}), but both now sum over half of the unit cells, see Fig.\ref{simple2D}(b). Just as before, the summation can be split in a sum over the even and odd unit cells, which for $H_{K1}$ reads
\begin{align}
\nonumber  H_{K1}=&-t\sum_{m,\sigma;\alpha\in\{x,y\}}\left[\hat{\psi}^{s\dagger}_{\sigma,2m}\hat{\psi}^{\alpha}_{\sigma,2m}+\hat{\psi}^{s\dagger}_{\sigma,2m+1}\hat{\psi}^{\alpha}_{\sigma,2m+1}\right.\\
&\left.-\hat{\psi}^{\alpha\dagger}_{\sigma,2m+\alpha}\hat{\psi}^s_{\sigma,2m}-\hat{\psi}^{\alpha\dagger}_{\sigma,(2m+1)+\alpha}\hat{\psi}^s_{\sigma,2m+1}+\text{h.c.}\right] \nonumber.
\end{align}
After using the same transformation for the fields as in the one-dimensional case, Eq.(\ref{evenoddtransformation}), the Hamiltonian reads
\begin{align}
\nonumber  H_{K1}&=-t\sum_{m,\sigma;\alpha\in\{x,y\}}\left[\tilde{\psi}^{s\dagger}_{\sigma,2m}\tilde{\psi}^{\alpha}_{\sigma,2m}+\tilde{\psi}^{s\dagger}_{\sigma,2m+1}\tilde{\psi}^{\alpha}_{\sigma,2m+1}\right.\\
&\left.+\tilde{\psi}^{\alpha\dagger}_{\sigma,2m+\alpha}\tilde{\psi}^s_{\sigma,2m}+\tilde{\psi}^{\alpha\dagger}_{\sigma,(2m+1)+\alpha}\tilde{\psi}^s_{\sigma,2m+1}+\text{h.c.}\right]\nonumber \\
&=-t\sum_{n',\sigma;\alpha\in\{x,y\}}\left[\tilde{\psi}^{s\dagger}_{\sigma,n'}\tilde{\psi}^{\alpha}_{\sigma,n'}+\tilde{\psi}^{\alpha\dagger}_{\sigma,n'+\alpha}\tilde{\psi}^s_{\sigma,n'}+\text{h.c.}\right]\nonumber,
\end{align}
where the summation over $n'$ runs over half of the values that $n$ runs over in the full 2D Hamiltonian, Eq.(\ref{2dspHamiltonianx}). The above Hamiltonian now describes particles hopping in a two-dimensional Lieb lattice with $s$-orbital sites only. The same transformation can be made for $H_{K2}$.
As in the one-dimensional case, the dispersions describing particles in the lattice with both $s$ and $p$ orbitals are of sine form, see Eq.(\ref{2dspHamiltoniank}), while particles in the Lieb lattice with only $s$ orbitals have a cosine dispersion. 
The Lieb lattice has been studied in the context of cold gases as well \cite{shen_single_2010,goldman_topological_2011}, where interesting phenomena result from the lattice exhibiting flat dispersions, so-called flat bands. Recently, Bose-Einstein condensation of atoms in a Lieb lattice was experimentally studied by  Taie {\it et al.}~\cite{taie_matter-wave_2015}.

\section{Pairing}
\label{sec:pairing}
In the previous section we studied particles in bipartite lattices and looked in detail at the effect of the different orbitals on the hopping. To explore the role of parity further, in this section we include on-site interactions between the fermionic particles in the one-dimensional lattice. We specifically look at pairing instabilities that can arise due to attractive interactions and study the different  superfluid phases that occur in this system. This section is meant as an instructive example, since the notion of a superfluid in one dimension is more complicated than the definition we use here.

The interaction Hamiltonian reads
\begin{align}
H_I^{1D}=\sum_n \left(U_0\hat{\psi}_{\uparrow,n}^{s\dagger}\hat{\psi}_{\downarrow,n}^{s\dagger}
\hat{\psi}_{\downarrow,n}^{s}\hat{\psi}_{\uparrow,n}^{s}+U_1\hat{\psi}_{\uparrow,n}^{p\dagger}
\hat{\psi}_{\downarrow,n}^{p\dagger}\hat{\psi}_{\downarrow,n}^{p}\hat{\psi}_{\uparrow,n}^{p}\right),
\end{align}
where we consider attractive interactions, $U_{0,1}<0$, which in general can be different. We use a mean-field approximation and include Cooper pairs $\Delta_{0}=U_{0}\langle\hat{\psi}_{\downarrow,n}^{s}\hat{\psi}_{\uparrow,n}^{s}\rangle$ and $\Delta_{1}=U_{1}\langle\hat{\psi}_{\downarrow,n}^{p}\hat{\psi}_{\uparrow,n}^{p}\rangle$, such that the total Hamiltonian reads
\begin{align}
\nonumber H^{1D}&=H_K^{1D}+H_I^{1D}\\
&=-\frac{|\Delta_0|^2}{U_0}-\frac{|\Delta_1|^2}{U_1}-2\mu_\downarrow+\sum_k\hat{\Psi}_k^\dagger\mathbb{H}_{BCS}\hat{\Psi}_k,
\label{totalHamiltonian}
\end{align}
where $\mu_\sigma$ is the chemical potential for fermions in spin state
$|\sigma\rangle$ and where the matrix $\mathbb{H}_{BCS}$ in the Nambu basis with $\hat{\Psi}^\dagger_{k}=(\hat{\psi}_{k\uparrow}^{s\dagger},\hat{\psi}_{-k \downarrow}^{s},\hat{\psi}_{k \uparrow}^{p\dagger},\hat{\psi}_{-k \downarrow}^{p})$
becomes
\begin{equation}
\mathbb{H}_{BCS}=
\begin{pmatrix}\label{hopmatrix}
-(\mu+h)&\Delta_0&-\text it\varepsilon_{k}&0\\
\Delta_0^*&\mu-h&0&\text it\varepsilon_{k}\\
\text it\varepsilon_{k}&0&-(\mu+h)&\Delta_1\\
0&-\text it\varepsilon_{k}&\Delta_1^*&\mu-h
\end{pmatrix},
\end{equation}
with the average chemical potential $\mu=(\mu_\uparrow+\mu_\downarrow)/2$ and where the possibility of having an imbalance in the population of the two spin components is included via a chemical potential difference $h=(\mu_\uparrow-\mu_\downarrow)/2$. For the spin-down sector we used $\varepsilon_{-k}=\sin(-k)=-\varepsilon_{k}$ for the dispersions in the hopping Hamiltonian Eq.(\ref{1dspHamiltoniank}) and we set $d=1$.
In order to write the total Hamiltonian using matrix multiplication, the spin-down fields have been interchanged. In the usual BCS theory this would result in extra terms $\varepsilon_k-\mu_\downarrow$ \cite{_ultracold_2008}, whereas here only the $2\mu_\downarrow$ in Eq.(\ref{totalHamiltonian}) stems from interchanging fermionic fields. This is due to the alternating signs for the hoppings, meaning that the two dispersions coming from interchanging spin-down fields cancel each other.
It is assumed in the above Hamiltonian that there is no energy offset between the $A$ and $B$ lattice sites.

\begin{figure}
\includegraphics[width=0.8\columnwidth]{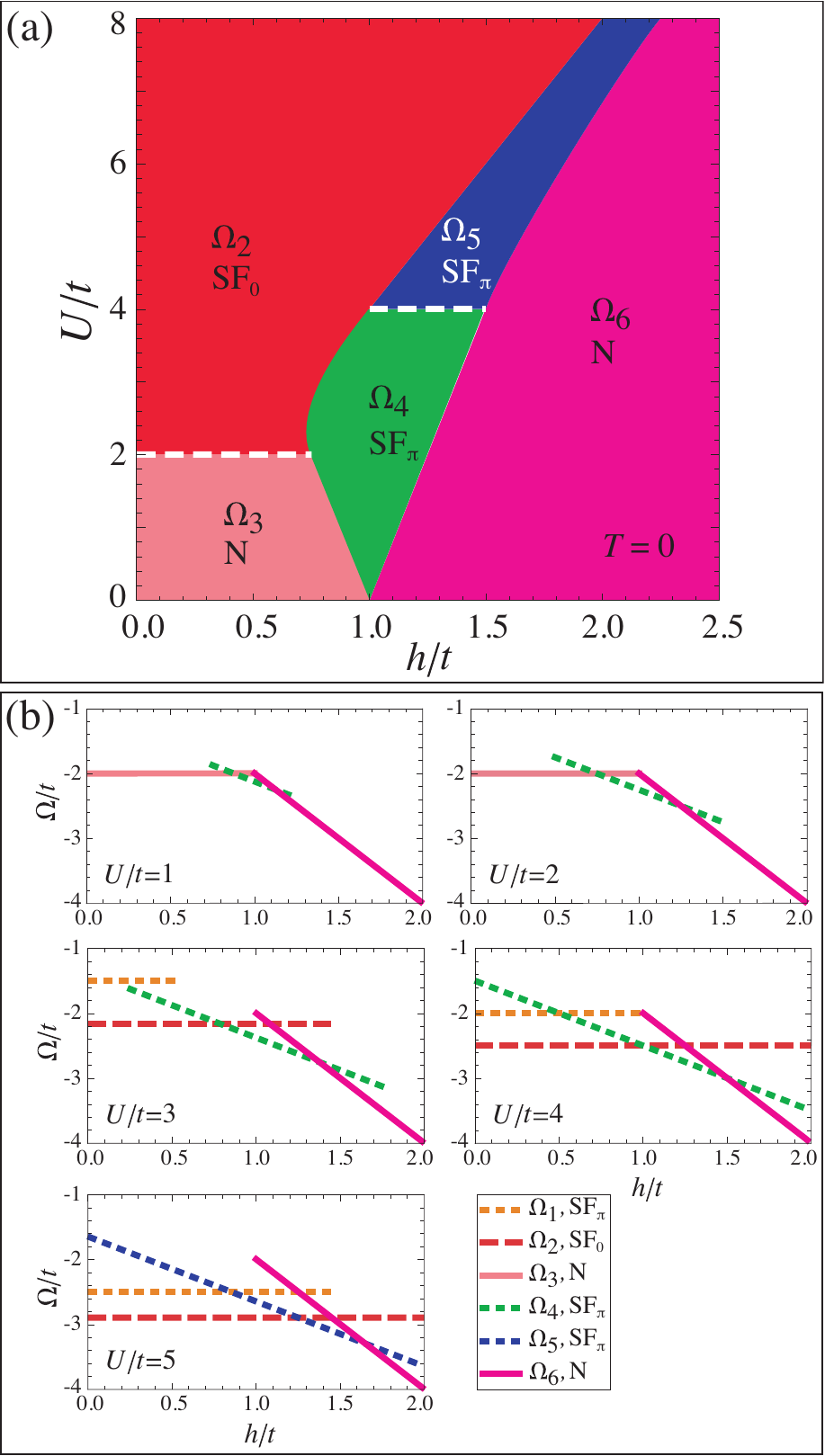}
\caption{ (a) Zero temperature phase diagram. Dashed lines denote continuous phase transitions, all other transitions are first order.
(b) Local minima of $\Omega^{1D}$ as functions of $h$ at different values of $U$. The different curves represent $\Omega_1-\Omega_6$ in Tab.~\ref{minima}, where the color coding is the same as in a). The dashed, dotted, and solid curves are for the cases of SF$_0$, SF$_\pi$, and normal phases, respectively.
}
\label{phasediag_simp}
\end{figure}

To understand the different phases that can occur in this system, we first neglect the momentum dependencies of the particle dispersions by setting them all equal to one, $\varepsilon_{k}=1$, for simplicity. By diagonalizing the above Hamiltonian the four quasiparticle dispersions $\hbar\omega_i$ are obtained, which apart from the usual $|\Delta_{0}|^2$ and  $|\Delta_{1}|^2$ terms, now also contain mixed terms, such as $\Delta_0\Delta_1$. Consequently, the partition function can be obtained $Z=\text{Tr}[\exp(-\beta H)]$, from which in turn the thermodynamic potential can be calculated $\Omega=-\ln Z/\beta$, where $\beta=1/k_BT$ is the inverse thermal energy, with $k_B$ Boltzmann's constant \cite{_ultracold_2008}. For our system the thermodynamic potential reads
\begin{align}
\nonumber \Omega^{1D}(\Delta_0,\Delta_1)=-\frac{|\Delta_0|^2}{U_0}-\frac{|\Delta_1|^2}{U_1}-2\mu_\downarrow\\
-\frac1\beta\sum_{i=1}^4\ln\left(1+e^{-\beta\hbar\omega_i}\right).
\label{thermpot}
\end{align}
We now minimize the thermodynamic potential $\Omega^{1D}$ with respect to the two pairing fields $\Delta_0$ and $\Delta_1$ at half filling and zero temperature $T=0$. A global minimum at $\Delta_0=\Delta_1=0$ corresponds to a phase without Cooper pairs, which is the normal phase, whereas a global minimum of $\Omega^{1D}$ at nonzero values for the pairing fields corresponds to a superfluid phase.
We take the interactions at the two sublattices to be equal, $U_0=U_1=U$ and map out the phase diagram as a function of the interaction strength $U/t$ and chemical potential difference $h/t$, see Fig.\ref{phasediag_simp}(a). Without imbalance and interactions the system is in the normal state ($\Omega_3$). For a large enough interaction, but still without a population imbalance, the thermodynamic potential is minimized by nonzero and equal pairing fields $\Delta_0=\Delta_1\neq 0$, which we refer to as the SF$_0$ phase ($\Omega_2$). This in contrast to the so-called $\pi$-phase, which we call here the SF$_\pi$ phase, where $\Delta_0$ and $\Delta_1$ have opposite sign. For large enough imbalance $h$ the thermodynamic potential is indeed minimized by nonzero $\Delta_0$ and $\Delta_1$ having opposite signs and the ground state of the system is the SF$_\pi$ phase. We find both an SF$_\pi$ phase where the pairing fields have equal magnitude ($\Omega_4$), $|\Delta_0|=|\Delta_1|$, and an SF$_\pi$ phase with unequal pairing fields ($\Omega_5$). For even larger imbalances $h$ the system enters the normal state again ($\Omega_6$).
We find that most of the above phase transitions take place with the order parameters changing discontinuously, suggesting first order phase transitions. The exceptions are between $\Omega_2$ and $\Omega_3$ at $U=2t$, and between $\Omega_4$ and $\Omega_5$ at $U=4t$, where the pairing fields change continuously.
In the above mentioned SF$_0$ phase the two pairing fields $\Delta_0$ and $\Delta_1$ take the same value, which means a constant total pairing field through out the lattice. This corresponds to a homogeneous superfluid phase, like in the usual BCS theory. Interestingly, in the SF$_\pi$ phases the pairing fields take different values on the different sublattices and the corresponding phase is not a homogeneous superfluid phase.

Because we neglected the momentum dependencies in the Hamiltonian in Eq.(\ref{hopmatrix}) and consequently ended up with a thermodynamic potential without momentum integrals it is possible to even find analytic expressions for the minima of the thermodynamic potential $\Omega$ and the pairing fields, $\Delta_0$ and $\Delta_1$, for which these minima are acquired. For half filling, we list all possible local minima and the phase they correspond to in Tab.~\ref{minima}, together with the pairing fields.
In Fig.~\ref{phasediag_simp}(b), we demonstrate the evolution of these local minima with the chemical potential difference $h$ for different interaction strengths $U$. From this comparison the global minimum can be identified, which is how the phase diagram in Fig.~\ref{phasediag_simp}(a) was obtained. 
Also listed in Tab.\ref{minima} for each phase are the polarizations $P$, where the polarization is the difference in densities between the two spin species divided by the total density, $P=(n_\uparrow-n_\downarrow)/(n_\uparrow+n_\downarrow)$. The spin component densities can be calculated from the thermodynamic potential, $n_\sigma=-\partial\Omega/\partial\mu_\sigma$.

\begin{table*}[t]
\begin{tabular}{|c|c|c|c|c|c|}
\hline
$\Omega_{1}$ & $\Omega_{2}$ & $\Omega_{3}$ & $\Omega_{4}$ & $\Omega_{5}$ & $\Omega_{6}$\\
\hline
SF$_\pi$(metastable) & SF$_0$ ($P=0$) & N ($P=0$) & \multicolumn{2}{c|}{SF$_\pi$ ($P=1/2$)} & N ($P=1$)\\
\hline
$0\leq h\leq U/2-t$ & $0\leq h\leq U/2$ & $0\leq h\leq t$ & $|h-t|\leq U/4$ & $0\leq h\leq U/2$ & $0<t\leq h$\\
$0<t<U/2$ & $0<t<U/2$ & $0<U/2\leq t$ & $0<U/4\leq t$ & $0<t<U/4$ & $0<U$\\
\hline
$\Delta_0=U/2$ & $\Delta_0=\sqrt{U^2/4-t^2}$ & $\Delta_0=0$ & $\Delta_0=U/4$ & $\Delta_0=U/4+\sqrt{U^2/16-t^2}$ & $\Delta_0=0$\\
$\Delta_1=-U/2$ & $\Delta_1=\sqrt{U^2/4-t^2}$ & $\Delta_1=0$ & $\Delta_1=-U/4$ & $\Delta_1=-U/4+\sqrt{U^2/16-t^2}$ & $\Delta_1=0$\\
\hline
$\Omega=-U/2$ & $\Omega=-2t^2/U-U/2$ & $\Omega=-2t$ & $\Omega=-h-t-U/8$ & $\Omega=-h-2t^2/U-U/4$ & $\Omega=-2h$\\
\hline
\end{tabular}
\caption{Possible local minima of the thermodynamic potential $\Omega^{1D}$ in Eq.(\ref{thermpot}) and their corresponding conditions at half filling, $\mu=0$, for equal interactions $U_0=U_1=U$. The polarization $P$ is also shown for each phase.
}\label{minima}
\end{table*}

\subsection{Connection between the inhomogeneous SF$_\pi$ and LO phase}
Thus, we find from the above simplified, one-dimensional model with alternating signs for the hopping parameter, corresponding to a lattice with alternating $s$ and $p$ orbital sites, that in the presence of an imbalance between the two spin components the ground state of the system can be formed by an inhomogeneous superfluid phase, the SF$_\pi$ phase. Namely, in contrast to a homogeneous superfluid phase, the SF$_0$ phase, where the pairing field is constant throughout the system, in the SF$_\pi$ phase the pairing field changes with position in the lattice.

We can compare the SF$_\pi$ phase to another inhomogeneous superfluid phase, the so-called Larkin-Ovchinnikov (LO) phase \cite{larkin_1965}. In an LO superfluid the order parameter is also taken to be position dependent and specifically to be a cosine, where the wave vector $q$ is left as a free parameter
\begin{align*}
\Delta(x)=\Delta_\text{LO}\cos(q x).
\end{align*}
This Cooper pair ansatz results in a thermodynamic potential that depends both on the pairing field amplitude $\Delta_\text{LO}$ and the wave vector $q$, $\Omega(|\Delta_\text{LO}|,q)$ \cite{baarsma_inhomogeneous_2013}. Physically, the above Cooper pair ansatz corresponds to pairs formed by two fermions with different momenta, e.g., $\hat\psi_{k\uparrow}$ and $\hat\psi_{q-k\downarrow}$. The order parameter wave vector is equal to the net momentum $q$ of the pairs and its wavelength is thus inversely proportional to it, $\lambda_\text{LO}=2\pi/q$.

\begin{figure}
\includegraphics[width=0.8\columnwidth]{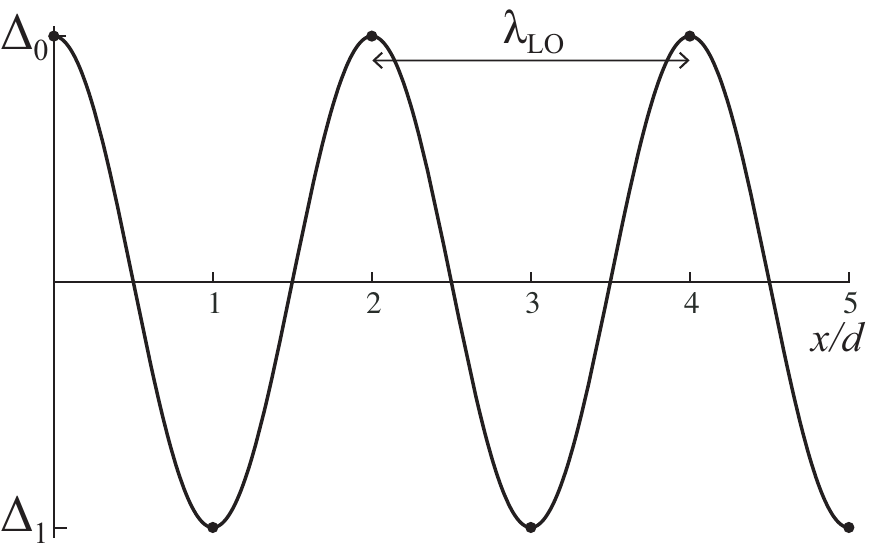}
\caption{LO order parameter and the SF$_\pi$ phase order parameters. For $\Omega_4$ the axes origin is at zero, whereas for $\Omega_5$ it is at some nonzero value.}\label{FFLO}
\end{figure}

Now, if the LO wavelength is twice the lattice spacing, $\lambda_\text{LO}=2d$, the LO phase in a lattice strongly resembles the SF$_\pi$ phase we find, where the pairing fields $\Delta_0$ and $\Delta_1$ only differ in sign ($\Omega_4$), see Fig.\ref{FFLO}. Namely, the LO order parameter then takes the same value $\Delta_\text{LO}$, but with opposite sign, on neighboring sites. The SF$_\pi$ phase where the pairing fields also have a different magnitude ($\Omega_5$) can be viewed as a combination of a constant and a standing wave order parameter.
In both cases, the SF$_\pi$ phase corresponds to LO Cooper pairs with a net momentum of $q=\pi/d$, such as $\hat\psi_{k\uparrow}$ pairing with $\hat\psi_{\pi/d-k\downarrow}$. The reason we can find this LO-like superfluid phase, without taking it into account explicitly is because in a lattice the above pair corresponds to two particles with the same lattice momentum, $k$ and $-k$, since $\pi/d$ is the size of the Brillouin zone for a lattice where a unit cell contains two sites.

It is possible that if a full LO ansatz is taken into account for this system, a  standing wave with a different wavelength $\lambda_\text{LO}$ is found to be the ground state of the system. However, the general statement remains true, that a spin imbalance can result in inhomogeneous superfluid phase.

Next we proceed to study a two-dimensional lattice, where there are two different $p$ orbitals on the $B$ sublattice, and we include the momentum dependence of the dispersions. In that case, taking a full LO ansatz into account would be more involved. Interestingly, we are still able to find inhomogeneous superfluid phases in a rather simple manner, via the possibility of the SF$_\pi$ phase.

\section{Experimentally realizable systems with hybridized orbitals}
\label{sec:exprealization}
In order to demonstrate how the phases revealed in the simple model can be observed experimentally, here we introduce a concrete two-dimensional lattice which has been realized recently by the group of Hemmerich \cite{wirth_evidence_2011} for ultracold bosonic atoms. In their experiment, a checkerboard lattice is created by two sets of orthogonal laser beams with shallow sites on one sublattice ($A$) and deeper sites on the other sublattice ($B$). By proper tuning of the relative depths of the two sublattices a system can be created where the lowest energy level of the $A$ sites (the $s$ band) are in resonance with the first energy level of the $B$ sites (the $p_x$ and $p_y$ bands). On all sites of the $B$ sublattice the $s$ bands are fully occupied, such that particles in the $p_x$ or $p_y$ band can not relax to the lowest band.
We study the possibility of superfluid phases for a two-component Fermi gas with a population imbalance loaded into such a lattice. To this end, we use a Hamiltonian that includes hopping between nearest neighboring sites and attractive on-site interactions. We start by considering the full lattice potential and arrive at an effective Hamiltonian by using a tight binding approximation.

\subsection{Tight Binding Approximation}\label{tightbinding}
The lattice potential used in experiment can be described by
\begin{equation}\label{potential}
V(x,y)=-V_0|\cos(k_0x)+e^{i\theta}\cos(k_0y)|^2,
\end{equation}
where $V_0$ is the average potential depth and $k_0$ is the wave vector determining the lattice spacing, $k_0=\pi/d$. In the following we take the recoil energy $E_R=\hbar^2k_0^2/(2m)$ as the unit of energy, and the distance between two adjacent minima of $A$ and $B$ sites $d$ as the unit of length.

If the lattice potential $V_0$ is large enough, a tight binding model can well describe the properties of the system. In this approximation, the traps at shallower $A$ and deeper $B$ sites can be taken as harmonic potentials by expanding the lattice potential around each site as
\begin{align}
\nonumber V^A&\approx-2V_0(1+\cos\theta)+V_0k_0^2(1+\cos\theta)(x^2+y^2)\\
\nonumber&\equiv E_{0}^A+\frac12\hbar\omega_A^2(x^2+y^2),\\
\nonumber V^B&\approx-2V_0(1-\cos\theta)+V_0k_0^2(1-\cos\theta)(x^2+y^2)\\
\nonumber&\equiv E_{0}^B+\frac12\hbar\omega_B^2(x^2+y^2),
\end{align}
where the energy levels of the oscillators are $E_n^{A,B}=E_0^{A,B}+\hbar\omega_{A,B}(n+1/2)$.
The degeneracy of the harmonic oscillator energy levels in two dimensions is $n+1$ with corresponding parity $(-1)^n$. Based on these energy levels, the $s$-band at the shallow lattice sites $E_0^A$ is in resonance with the $p_x$ and $p_y$ bands on the deeper lattice sites $E_1^B$ when $E_0^A+\hbar\omega^A/2=E_0^B+3\hbar\omega^B/2$, which gives a relation between the lattice depth $V_0$ and the phase $\theta$ in Eq.(\ref{potential}).
By numerically calculating the energy bands from the full lattice potential, we find for a lattice depth $V_0=10$ that the energy bands are in resonance for $\theta\approx0.556\pi$, while from the harmonic oscillator energy levels one finds $\theta\approx0.560\pi$.
This difference is small, which ensures that the tight binding is a good approximation.

In the following, we focus on the above mentioned three orbitals, $s$ on the $A$ sublattice and $p^x$ and $p^y$ on the $B$ sublattice in resonance with each other. The chemical potentials of the system are chosen such that at low enough temperatures other bands are either fully occupied or empty and therefore play no role in our present study on superfluidity.

\begin{figure}
\includegraphics[width=\columnwidth]{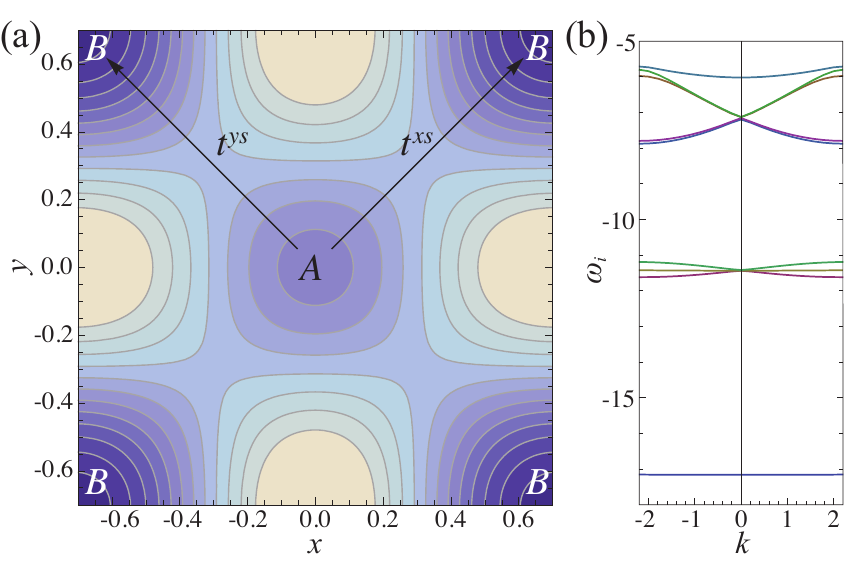}
\caption{(a) Lattice potential in Eq.~(\ref{potential}) with $V_0=10$ and $\theta\approx0.556\pi$ in one unit cell, with the shallower $A$ site at the origin and the deeper $B$ sites at four corners. The bottom of the potential at $B$ sites is $-24.67$, while it is $-15.33$ for $A$ sites. The two arrows indicate nearest neighbor hopping. (b) Corresponding dispersions of single particle states along $k_x=k_y$}\label{potential&disper}
\end{figure}

The lattice potential in one unit cell and the corresponding band structure obtained by solving numerically the single-particle Schr\"odinger equation are shown in Fig.~\ref{potential&disper}. There, the lowest dispersion corresponds to the lowest energy band in the $B$ sublattice. The first three bands above this band are the bands of interest, $p^x$, $p^y$, and $s$, hybridized by the hopping. The even higher dispersions correspond to even higher energy bands in the lattice. It can be seen that the three bands of interest are well separated from the other bands and considering their hybridization, the Hamiltonian term $H^{AB}$ corresponding to the nearest-neighbour hopping can indeed be described by Eq.~(\ref{2dspHamiltonianx}).

If we also include next nearest neighbor hopping, we can write down a total single-particle Hamiltonian $H_0$ in the basis $\hat{\Psi}_{\bf k}^\dagger=(\hat{\psi}^{s\dagger}_\mathbf{k},\hat{\psi}^{x\dagger}_\mathbf{k},\hat{\psi}^{y\dagger}_\mathbf{k})$,
\begin{align}
H_0=\sum_{\bf k}\hat{\Psi}_{\bf k}^\dagger \mathbb{H}_K \hat{\Psi}_{\bf k},
\end{align}
with the matrix
\begin{widetext}
\begin{equation}\label{hybrid_moment}
\mathbb{H}_K=\begin{pmatrix}
\epsilon_0^A+4t^{ss}\cos k_x\cos k_y&2\text it^{xs}\sin k_x&2\text it^{ys}\sin k_y\\
-2\text it^{xs}\sin k_x&\epsilon_0^B+2t^{xx}\cos k_x\cos k_y&2t^{xy}\sin k_x\sin k_y\\
-2\text it^{ys}\sin k_y&2t^{xy}\sin k_x\sin k_y&\epsilon_0^B+2t^{yy}\cos k_x\cos k_y
\end{pmatrix},
\end{equation}
\end{widetext}
where the on-site energy offsets $\epsilon_0^{A,B}$ for $A$ and $B$ sites were added.

\begin{figure}
\includegraphics[width=0.6\columnwidth]{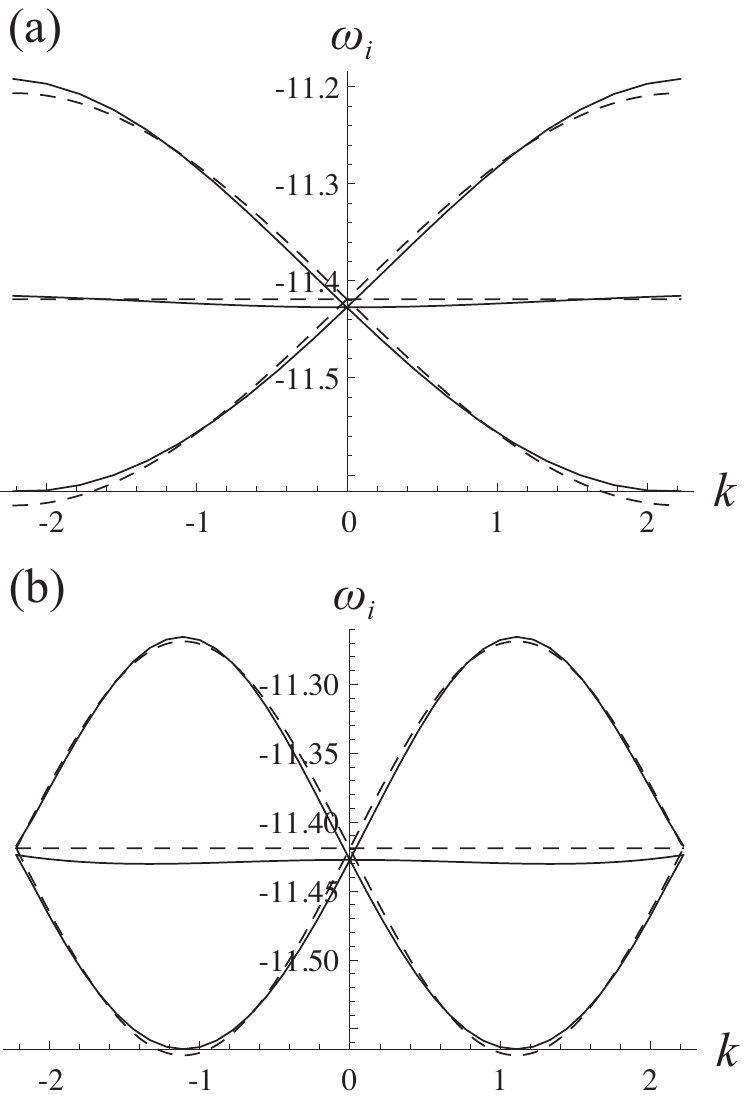}
\caption{Dispersions of the three hybridized states along (a) $k_x=k_y$ and (b) along $k_x$ with $k_y=0$, for the lattice with $V_0=10$ and $\theta\approx0.556\pi$. The solid curves are exact dispersions solved numerically, while the dashed curves are the fitted dispersions of the reduced Hamiltonian with only two parameters $\epsilon_0$ and $t$.
}\label{fitdisperls}
\end{figure}

The hopping coefficients and energy offsets can be obtained by fitting the dispersions obtained from diagonalizing the above Hamiltonian to the exact dispersions calculated numerically. As an example, for the hopping parameters we find $t^{xs}=t^{ys}\approx0.0747$ and for the energy offsets $\epsilon_0^A\approx-11.42$ and $\epsilon_0^B\approx-11.41$, in the case of lattice parameters $V_0=10$ and $\theta\approx0.556\pi$, whereas the next nearest neighbor hoppings are at least three orders of magnitude smaller.
Therefore, it is possible to use a reduced Hamiltonian with only nearest neighbor hoppings $t\equiv t^{xs}=t^{ys}$  and also, because $\epsilon_0^{A,B}$ are almost the same, they can be replaced by one parameter $\epsilon_0$. If we now fit the dispersions obtained from the reduced Hamiltonian with the exact dispersions, we find $t\approx0.0751$ and $\epsilon_0\approx-11.42$. The three dispersions of the new hybrid states are $E^1_\mathbf{k}=\epsilon_0$ and $E^{2,3}_\mathbf{k}=\epsilon_0\pm2t\sqrt{\sin^2k_x+\sin^2k_y}$, which reproduce the numerical results very well, see Fig.~\ref{fitdisperls}. In the following we will use the reduced  Hamiltonian.

Here, it is worth pointing out that band $E^1$ is exactly flat, i.e., dispersionless, which results from the linear combination of the $p^x$ and $p^y$ orbitals in the nearest-neighbor hopping approximation. Actually, the dispersions depicted in Fig.~\ref{fitdisperls} are the same as the dispersions in the Lieb lattice, which illustrates the mapping discussed in the previous section.
A final remark in this section is that the discussion above considers only the orbital degrees of freedom, and is valid for both fermionic spin species we consider.

\subsection{On-site Interactions}\label{pairinteraction}
To study the possibility of a pairing instability, like previously, we now include interactions and study the different  superfluid phases that can occur in this two-dimensional system.

We include an attractive $s$-wave contact interaction
\begin{align}
H_I=U\int d\mathbf{r}\hat\psi_\uparrow^\dagger(\mathbf{r})\hat\psi_\downarrow^\dagger(\mathbf{r})\hat\psi_\downarrow(\mathbf{r})\hat\psi_\uparrow(\mathbf{r}),
\end{align}
where the interaction strength $U<0$. By expanding the fermionic operators using the Wannier states the interaction Hamiltonian reads,
\begin{align}
&U \int d\mathbf{r}\hat\psi^\dagger_\uparrow(\mathbf{r})\hat\psi^\dagger_\downarrow(\mathbf{r})\hat\psi_\downarrow(\mathbf{r})\hat\psi_\uparrow(\mathbf{r})\nonumber\\
&\approx U\sum_{\mathbf{R},\{n_i\}}\int d\mathbf{r}w^*_{n_1}(\mathbf{r-R})w^*_{n_2}(\mathbf{r-R})\nonumber\\ &\qquad\qquad w_{n_3}(\mathbf{r-R})w_{n_4}(\mathbf{r-R})\hat\psi^{n_1\dagger}_{\mathbf{R}\uparrow}\hat\psi^{n_2\dagger}_{\mathbf{R}\downarrow}\hat\psi^{n_3}_{\mathbf{R}\downarrow}\hat\psi^{n_4}_{\mathbf{R}\uparrow},\nonumber\\
&=\sum_{\mathbf{R},\{n_i\}}U_{n_1n_2n_3n_4}\hat\psi^{n_1\dagger}_{\mathbf{R}\uparrow}\hat\psi^{n_2\dagger}_{\mathbf{R}\downarrow}\hat\psi^{n_3}_{\mathbf{R}\downarrow}\hat\psi^{n_4}_{\mathbf{R}\uparrow},
\end{align}
with $n_i$ denoting the $s$, $p_x$ and $p_y$ orbitals and $\bf R$ the position of the unit cell, and where we used the localizing property of the Wannier functions. The effective interaction coefficients $U_{n_1n_2n_3n_4}$ absorb the corresponding cross integrals of four Wannier functions and are independent of $\mathbf{R}$ since all unit cells are equivalent in an infinite lattice. 
We use the harmonic oscillator eigenstates as an approximation to the Wannier functions to calculate the interaction coefficients. We only need to consider combinations of the $s$ band and the neighboring $p_x$ and $p_y$ bands within one unit cell, since all other cross integrals are at least four orders of magnitude smaller and can therefore be neglected.
The results of the dominant interaction integrals are shown in Table \ref{integralres}, where it is used that the absolute value of a cross integral does not depend on the order of the Wannier functions

\begin{table}[t]
\begin{tabular}{|c | c | c|}
\hline
$U_0\equiv U_{ssss}$ & $U_1\equiv U_{xxxx}\text{, }U_{yyyy}$ & $U_2\equiv U_{xxyy}$\\
4.51& 4.04 & 1.35\\ \hline
\end{tabular}
\caption{Numerical values of $U_{n_1n_2n_3n_4}/U$ for the lattice with $V_0=10$ and $\theta\approx0.556\pi$.
}\label{integralres}
\end{table}

The effective coupling constants are defined $U_0\equiv U_{ssss}$, $U_1\equiv U_{xxxx}=U_{yyyy}$, and $U_2\equiv U_{xxyy}$, where in the harmonic approximation $U_1=3U_2$, independent of the lattice potential depth.
In the $U_2$ interaction terms, the four orbitals yield six different combinations, namely $\hat\psi^{x\dagger}_\uparrow\hat\psi^{y\dagger}_\downarrow\hat\psi^y_\downarrow\hat\psi^x_\uparrow$, $\hat\psi^{x\dagger}_\uparrow\hat\psi^{y\dagger}_\downarrow\hat\psi^x_\downarrow\hat\psi^y_\uparrow$, $\hat\psi^{x\dagger}_\uparrow\hat\psi^{x\dagger}_\downarrow\hat\psi^y_\downarrow\hat\psi^y_\uparrow$, and these terms with $x$ and $y$ interchanged, which are all included in our model.

Using a simple mean-field approximation we introduce the BCS order parameters $\Delta^{nm}\equiv \sum_\mathbf{k}U_{mnnm}\langle\hat\psi^n_{\mathbf{k}\downarrow}\hat\psi^m_{\mathbf{-k}\uparrow}\rangle$. Considering the lattice symmetry, we have three different pairing fields denoted as $\Delta_0=\Delta^{ss}$, $\Delta_1=\Delta^{xx}=\Delta^{yy}$, and $\Delta_2=\Delta^{xy}=\Delta^{yx}$. For example, for the $U_2\hat\psi^{x\dagger}_\uparrow\hat\psi^{x\dagger}_\downarrow\hat\psi^y_\downarrow\hat\psi^y_\uparrow$ interaction term the mean-field approximation is as follows
\begin{align}
\nonumber & U_2\hat\psi^{x\dagger}_\uparrow\hat\psi^{x\dagger}_\downarrow\hat\psi^y_\downarrow\hat\psi^y_\uparrow\\
\nonumber &\simeq U_2\langle\hat\psi^{x\dagger}_\uparrow\hat\psi^{x\dagger}_\downarrow\rangle\hat\psi^y_\downarrow\hat\psi^y_\uparrow+
U_2\langle\hat\psi^y_\downarrow\hat\psi^y_\uparrow\rangle\hat\psi^{x\dagger}_\uparrow\hat\psi^{x\dagger}_\downarrow\\
\nonumber & -U_2\langle\hat\psi^{x\dagger}_\uparrow\hat\psi^{x\dagger}_\downarrow\rangle\langle\hat\psi^y_\downarrow\hat\psi^y_\uparrow\rangle\\
\nonumber &=U_2\frac{\Delta_1^*}{U_1}\hat\psi^y_\downarrow\hat\psi^y_\uparrow+U_2\frac{\Delta_1}{U_1}\hat\psi^{x\dagger}_\downarrow\hat\psi^{x\dagger}_\uparrow-U_2\frac{|\Delta_1|^2}{U_1^2}\\
&=\frac{\Delta_1^*}{3}\hat\psi^y_\downarrow\hat\psi^y_\uparrow+\frac{\Delta_1}{3}\hat\psi^{x\dagger}_\downarrow\hat\psi^{x\dagger}_\uparrow-\frac{|\Delta_1|^2}{3U_1}.
\end{align}
For all other interaction terms the mean-field approximation is similar.

With the mean-field pairing, we understand that there are four interacting channels included in our Hamiltonian, namely $\hat\psi^{n\dagger}_\uparrow\hat\psi^{n\dagger}_\downarrow\hat\psi^n_\downarrow\hat\psi^n_\uparrow$ ($n=s,x,y$) counts intraband pairing, $\hat\psi^{x\dagger}_\uparrow\hat\psi^{x\dagger}_\downarrow\hat\psi^y_\downarrow\hat\psi^y_\uparrow$ yields interband pair tunneling, $\hat\psi^{x\dagger}_\uparrow\hat\psi^{y\dagger}_\downarrow\hat\psi^y_\downarrow\hat\psi^x_\uparrow$ results in interband pairing and $\hat\psi^{x\dagger}_\uparrow\hat\psi^{y\dagger}_\downarrow\hat\psi^x_\downarrow\hat\psi^y_\uparrow$ corresponds to spin exchange within interband pairs. However, as shown below, the last two interband pairing terms turn out to have no contribution.

\subsection{Full Hamiltonian}
\label{sec:fullhamiltonian}
Including the nearest-neighbour hopping and the pairing terms, as well as a population imbalance, the total mean-field Hamiltonian can be written with the Nambu basis $\hat\Psi^\dagger_\mathbf{k}=(\hat\psi^{s\dagger}_{\mathbf{k}\uparrow},\hat\psi^{x\dagger}_{\mathbf{k}\uparrow},\hat\psi^{y\dagger}_{\mathbf{k}\uparrow},\hat\psi^{s}_{-\mathbf{k}\downarrow},\hat\psi^{x}_{-\mathbf{k}\downarrow},\hat\psi^{y}_{-\mathbf{k}\downarrow})$,

\begin{align}\label{fullHamilton}
\nonumber\frac{H}{N}=&\sum_\mathbf{k}\left\{\hat{\Psi}^\dagger_\mathbf{k}\mathbb{H}_{BCS}\hat{\Psi}_\mathbf{k}
+3[\epsilon_0-(\mu-h)]\right\}\\
&-\frac{|\Delta_0|^2}{U_{0}}-\frac{8|\Delta_1|^2}{3U_1}-\frac{4|\Delta_2|^2}{U_2},
\end{align}
with the matrix
\begin{widetext}
\begin{align}\label{Hamiltonianmatrix}
\mathbb{H}_{BCS}=\begin{pmatrix}
\epsilon_0-\mu-h&2\text it\sin k_x&2\text it\sin k_y&\Delta_0&0&0\\
-2\text it\sin k_x&\epsilon_0-\mu-h&0&0&4\Delta_1/3&2\Delta_2\\
-2\text it\sin k_y&0&\epsilon_0-\mu-h&0&2\Delta_2&4\Delta_1/3\\
\Delta_0^{*}&0&0&-\epsilon_0+\mu-h&-2\text it\sin k_x&-2\text it\sin k_y\\
0&4\Delta_1^*/3&2\Delta_2^{*}&2\text it\sin k_x&-\epsilon_0+\mu-h&0\\
0&2\Delta_2^*&4\Delta_1^*/3&2\text it\sin k_y&0&-\epsilon_0+\mu-h
\end{pmatrix}.
\end{align}
\end{widetext}
Accordingly, the thermodynamic potential reads
\begin{align}\label{Omega}
\nonumber \Omega(\Delta_0,\Delta_1,\Delta_2)=&\frac{1}{\mathcal{V}}\sum_\mathbf{k}\bigg\{3[\epsilon_0-(\mu-h)]\\
&-\frac1\beta\sum_i\ln\left[1+e^{-\beta\omega_i(\mathbf{k})}\right]\bigg\}\nonumber\\ &-\frac{|\Delta_0|^2}{U_0\mathcal{V}}-\frac{8|\Delta_1|^2}{3U_1\mathcal{V}}-\frac{4|\Delta_2|^2}{U_2\mathcal{V}},
\end{align}
where $\mathcal{V}$ is the 2D volume of a unit cell, $\omega_i(\mathbf{k})$ are the six eigenvalues from the $6\times6$ matrix in Eq.~(\ref{Hamiltonianmatrix}), and the quasi-momentum summation is over the first Brillouin zone.

\section{Results}
\begin{figure}
\includegraphics[width=0.995\columnwidth]{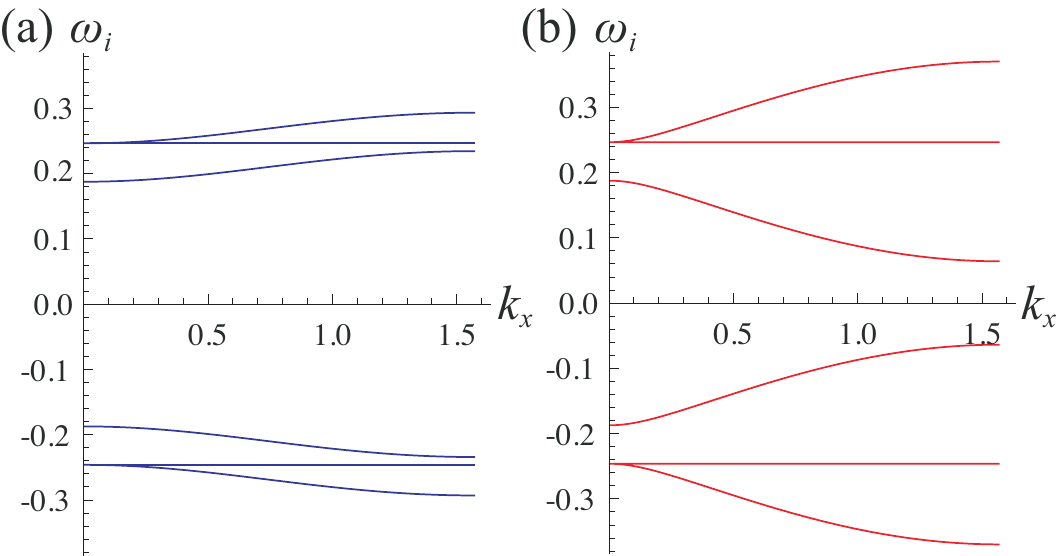}
\caption{Dispersions $\omega_i$ for (a) the SF$_0$ phase and for (b) the SF$_\pi$ phase as functions of $k_x$ at fixed $k_y=0$, with the parameters $t\approx0.075$ and $\epsilon_0=\mu\approx-11.42$, and lattice parameters $V_0=10$ and $\theta\approx0.556\pi$. In panel (a) $\Delta_0\approx0.187$ and $\Delta_1\approx0.185$, which minimize $\Omega$ at $h=0$, $U=0.10$ and $T=0.01$, while in panel (b) the sign of $\Delta_1$ is reversed to make a comparison between the SF$_0$ and SF$_\pi$ phases.}\label{dispersh}
\end{figure}
From the thermodynamic potential Eq.(\ref{Omega}) we can, like earlier, obtain phase diagrams for the 2D lattice with $s$ and $p$ orbital sites. We obtain phase diagrams as function of interaction $U$ and imbalance $h$ for different temperatures and lattice parameters by minimizing $\Omega$ with respect to the order parameters $\Delta_0$, $\Delta_1$ and $\Delta_2$. For the parameter regimes we considered, we find that $\Delta_2$ is always zero by numerically minimizing $\Omega$. Subsequently, we calculate the momentum distributions for the spin-particles for the different phases we find. But first we take a look at the dispersions. 

\subsection{Dispersions}
The thermodynamic potential is calculated from the eigenvalues of $\mathbb{H}_{BCS}$ in Eq.(\ref{Hamiltonianmatrix}), $\omega_i(\bf k)$, which in the normal state ($\Delta_0=\Delta_1=0$) correspond to the particle dispersions and in the case of a superfluid phase to the quasi-particle dispersions. As in the 1D case, we find superfluid phases with both $\Delta_0$ and $\Delta_1$ being nonzero, having either the same sign (SF$_0$) or the opposite sign (SF$_\pi$). In Fig.\ref{dispersh} the dispersions $\omega_i$ for the SF$_0$ phase, panel (a), and the SF$_\pi$ phase, panel (b), can be compared, where all parameters were chosen the same to calculate these figures and the only difference is an added minus sign to $\Delta_1$ for Fig.\ref{dispersh}(b). The flat dispersions are the same for the two cases.

The dispersions are shown for the population balanced system. In the presence of a population imbalance, i.e. $h>0$, the dispersions are shifted downwards(upwards) for the spin up(down) particles, which are then the majority(minority) particles.  Intuitively, it can then be understood from these dispersions that depending on the imbalance it is energetically more favourable to either occupy quasi-particle states corresponding to the SF$_0$ phase or to the SF$_\pi$ phase. However, to obtain the exact phase diagram of course the full thermodynamic potential should be minimized, which is what we do next.

\subsection{Phase diagrams}
\begin{figure*}
\includegraphics[width=0.67\columnwidth]{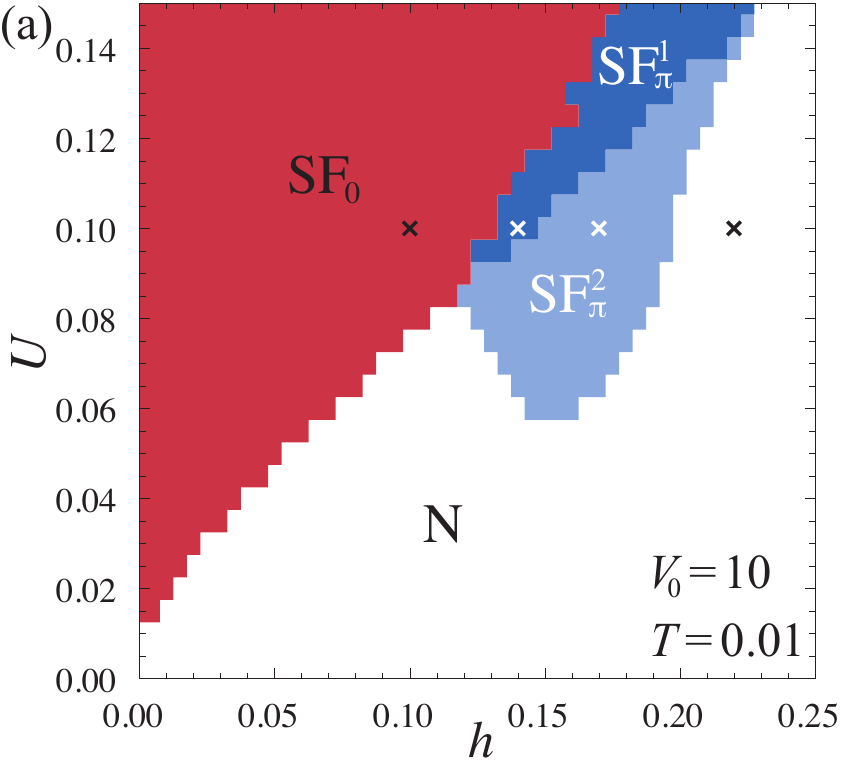}
\includegraphics[width=0.67\columnwidth]{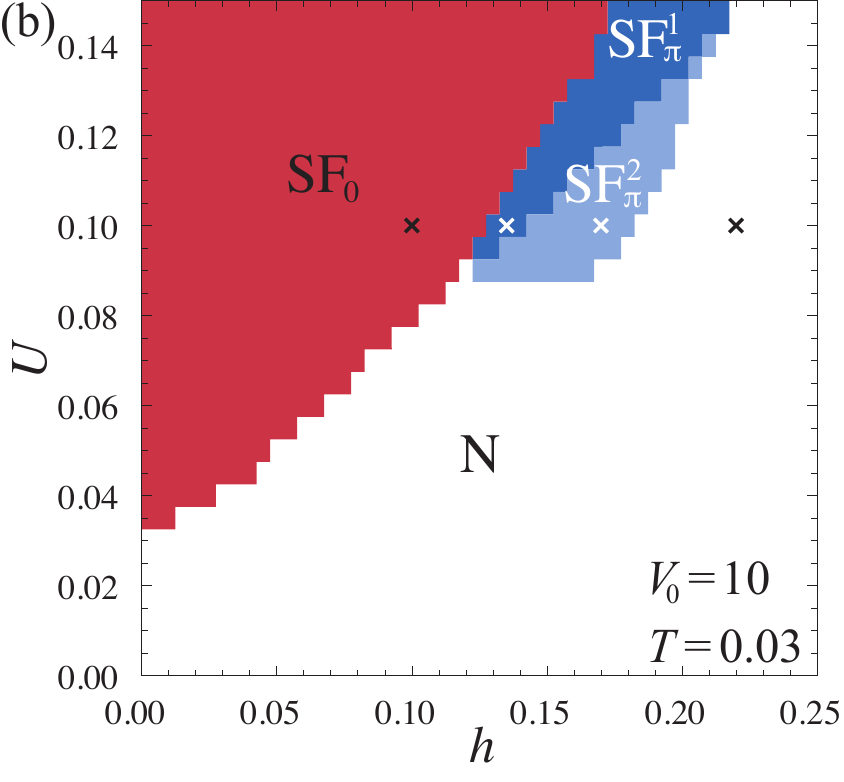}
\includegraphics[width=0.67\columnwidth]{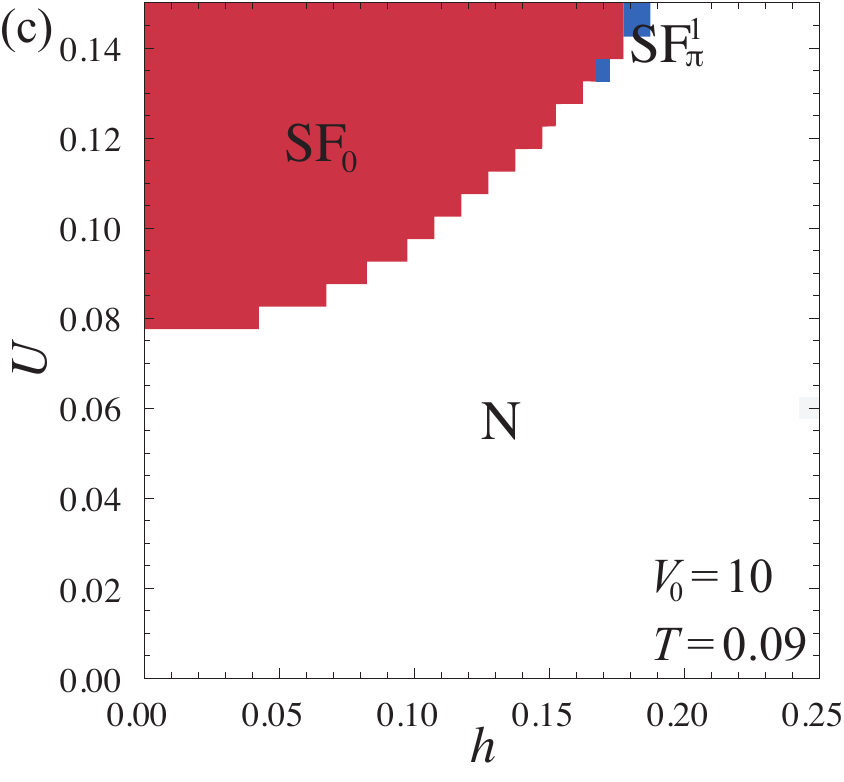}
\caption{Phase diagrams as functions of $h$ and $U$ for lattice parameters $V_0=10$ and $\theta\approx0.556\pi$ at different temperatures. In all phase diagrams the white region denotes the normal phase, red corresponds to the SF$_0$ and blue to the SF$_\pi$ phases. The crosses mark the values for which the momentum distributions are calculated in Fig.\ref{momentumU010T001} and Fig.\ref{momentumU010T003}.
}\label{phasediag_real}
\end{figure*}
We now present phase diagrams as functions of chemical potential difference $h$ and interaction strength $U$. Here, $U$ is the full interaction strength from which the effective interactions $U_0$ and $U_1$ are calculated and is different from the interaction coefficient used in the 1D case.
By minimizing $\Omega$ with fixed $\mu=\epsilon_0$, we find numerically that $\Delta_2$ always vanishes, while $\Delta_0$ and $\Delta_1$ have similar behaviour as we found for the simple 1D model. 

In Fig.\ref{phasediag_real} phase diagrams are shown for the lattice parameters $V_0=10$ and $\theta\approx 0.556\pi$ at different temperatures. White regions correspond to the normal phase, red to the SF$_0$ phase and (darker and lighter) blue corresponds to SF$_\pi$ phases. In contrast to the 1D case, here the SF$_\pi$ phase with $\Delta_0=-\Delta_1$ ($\Omega_4$) is missing or at least highly reduced. The SF$_\pi$ phase with $|\Delta_0|\neq|\Delta_1|$ ($\Omega_5$) splits into two phases, one with $|\Delta_0|<|\Delta_1|$ (SF$_\pi^1$) and one with $|\Delta_0|>|\Delta_1|$ (SF$_\pi^2$). The split of this $\Omega_5$ phase was to be expected, since now $U_0$ is not equal to $U_1$ and thereby the degeneracy between the two local minima of $\Omega$ is lifted.

We also observe that, with increasing temperature, the superfluid phases shrink towards the larger $U$ and smaller $h$ corner, with the SF$_\pi$ phase completely disappearing for high enough temperatures. 

\begin{figure}
\includegraphics[width=0.67\columnwidth]{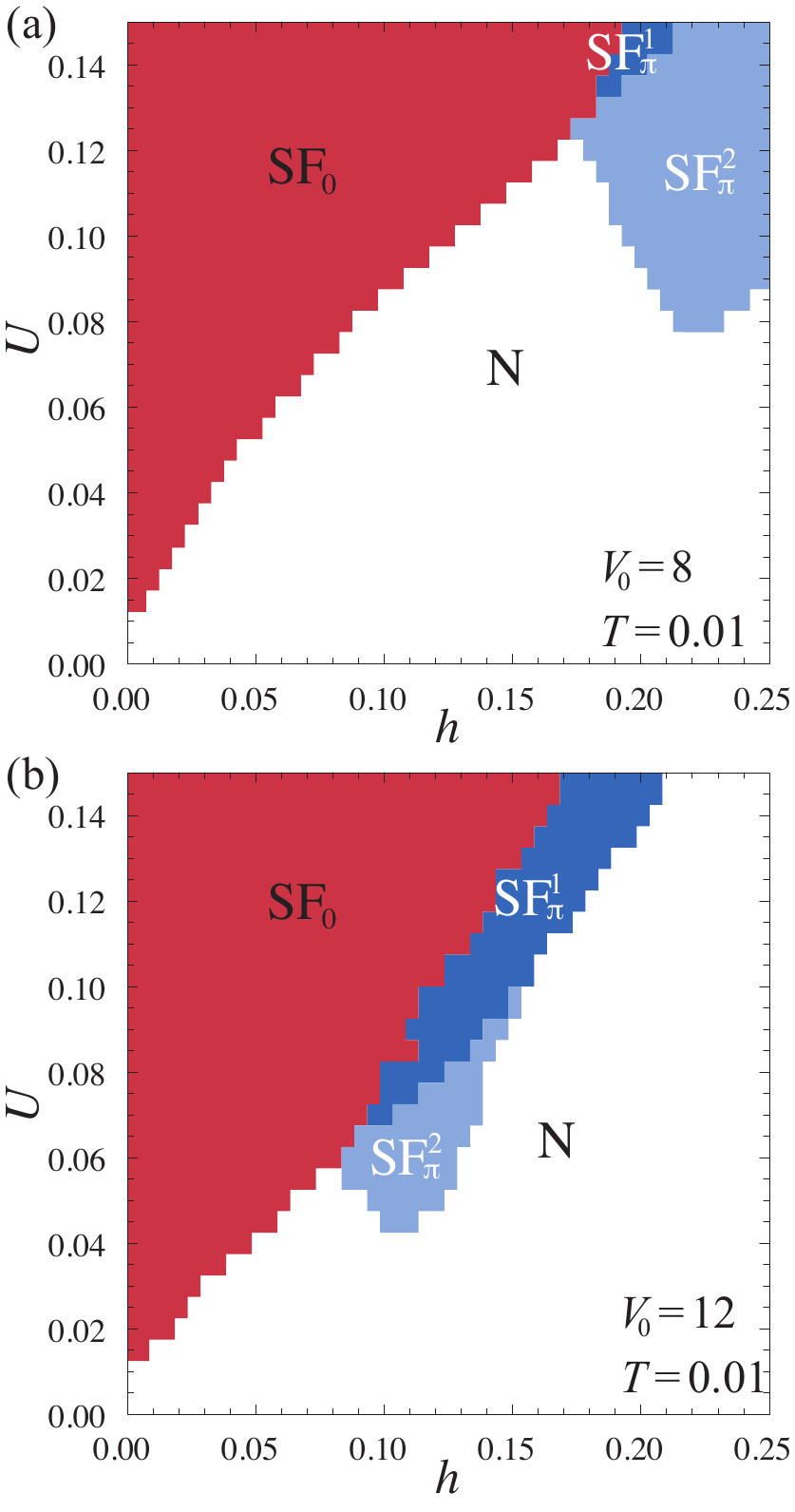}
\caption{Phase diagrams as functions of $h$ and $U$ for fixed temperature $T=0.01$ at different lattice parameters. A shallower lattice is considered in (a) with $V_0=8$, $\theta\approx 0.564\pi$, $t\approx 0.110$, $\epsilon_0\approx-8.44$, $U_0\approx 3.98U$ and $U_1\approx 3.65U$. A deeper lattice is used in (b) with $V_0=12$ and $\theta\approx 0.550\pi$, with $t\approx 0.0517$, $\epsilon_0\approx-14.5$, $U_0\approx 5U$ and $U_1\approx 4.39U$. As before, the white regions denote the normal phase (N), red corresponds to the SF$_0$ and blue to the SF$_\pi$ phases.}\label{phasediag_latt}
\end{figure}

Furthermore, to study the effect of different hopping parameters $t$, we change the lattice potential via $V_0$ and $\theta$, which can also be modified experimentally. In this way, the parameters $t$ and $\epsilon_0$ obtained from the fitting in the tight binding model, as well as the effective interactions $U_i$, are modified. We consider both a shallower lattice with $V_0=8$ and a deeper lattice with $V_0=12$, where the phase diagrams for these two cases are plotted in Fig.~\ref{phasediag_latt}(a) and (b) respectively. The lattice with $V_0=8$ corresponds to a larger hopping coefficient $t\approx0.110$ than previously, meaning that the other energy scales in the Hamiltonian, the interaction $U$ and the imbalance $h$, become effectively smaller. The result is that the same phases as before now occur for larger $U$ and $h$, which can be observed in the phase diagram Fig.~\ref{phasediag_latt}(a). The deeper lattice with $V_0=12$ corresponds to a smaller hopping $t\approx0.0517$ and we observe the opposite effect. The superfluid SF$_\pi$ phase region now shifts towards smaller $h$ and $U$.

\begin{figure*}
\includegraphics[width=0.24\textwidth]{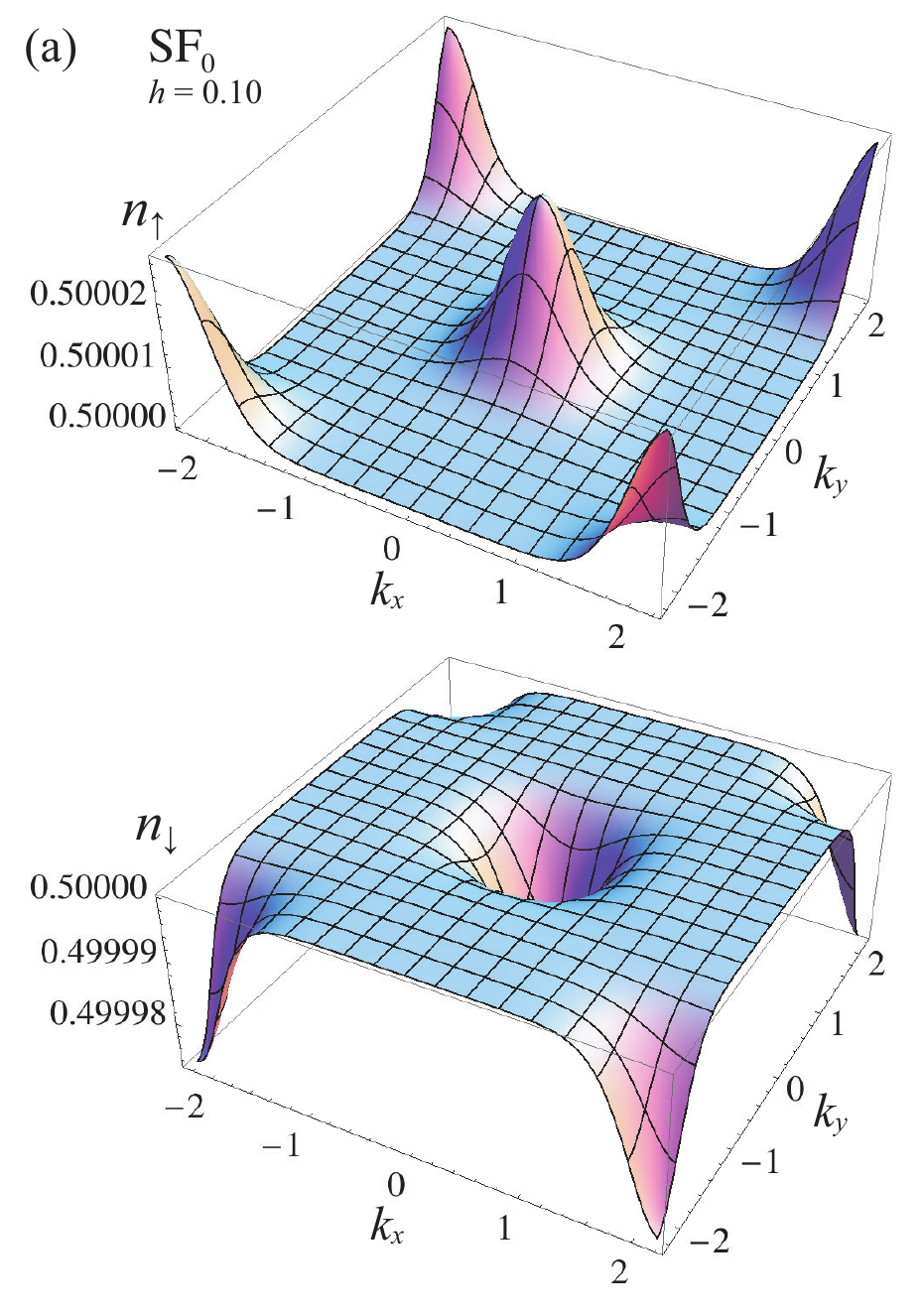}
\includegraphics[width=0.24\textwidth]{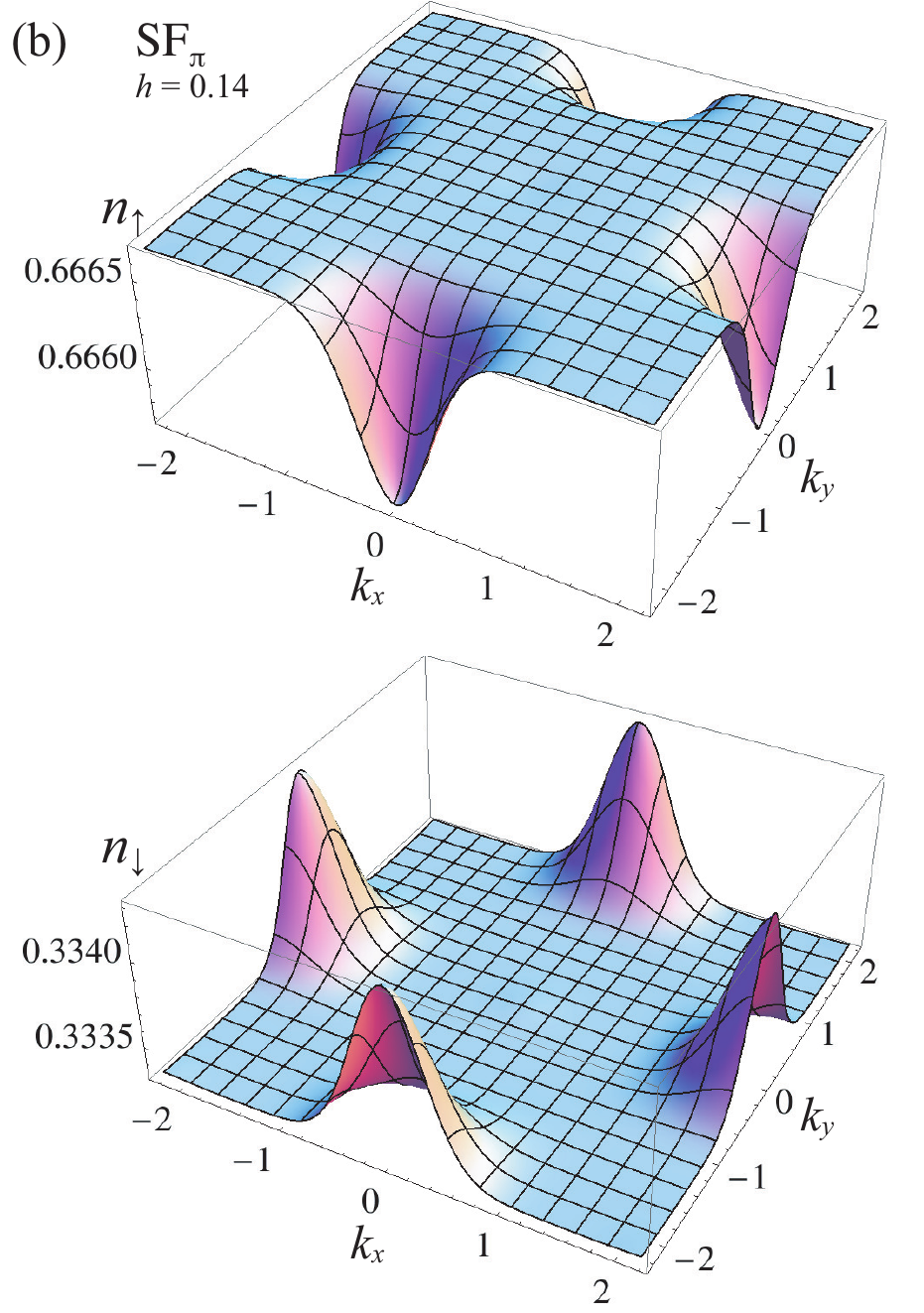}
\includegraphics[width=0.24\textwidth]{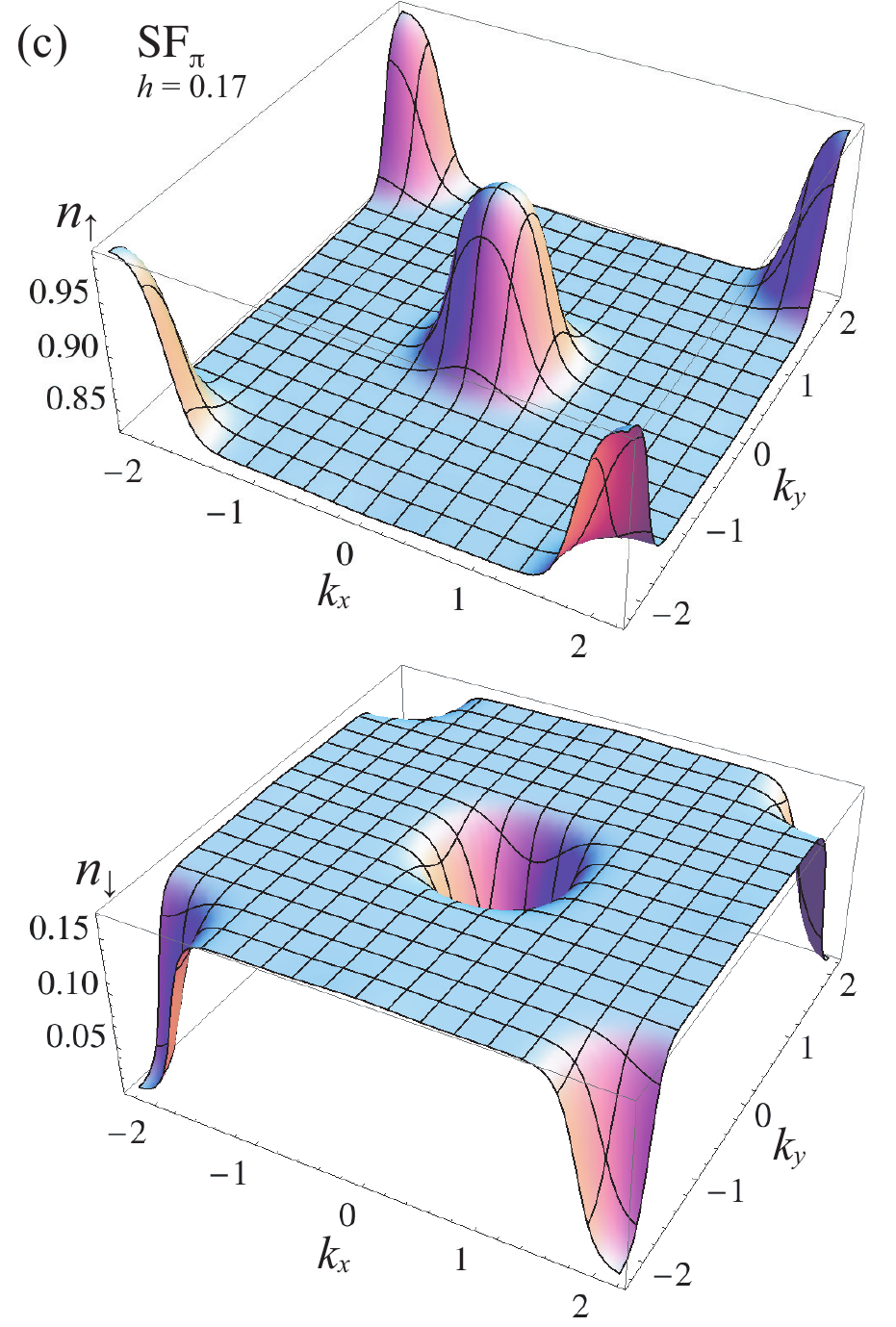}
\includegraphics[width=0.24\textwidth]{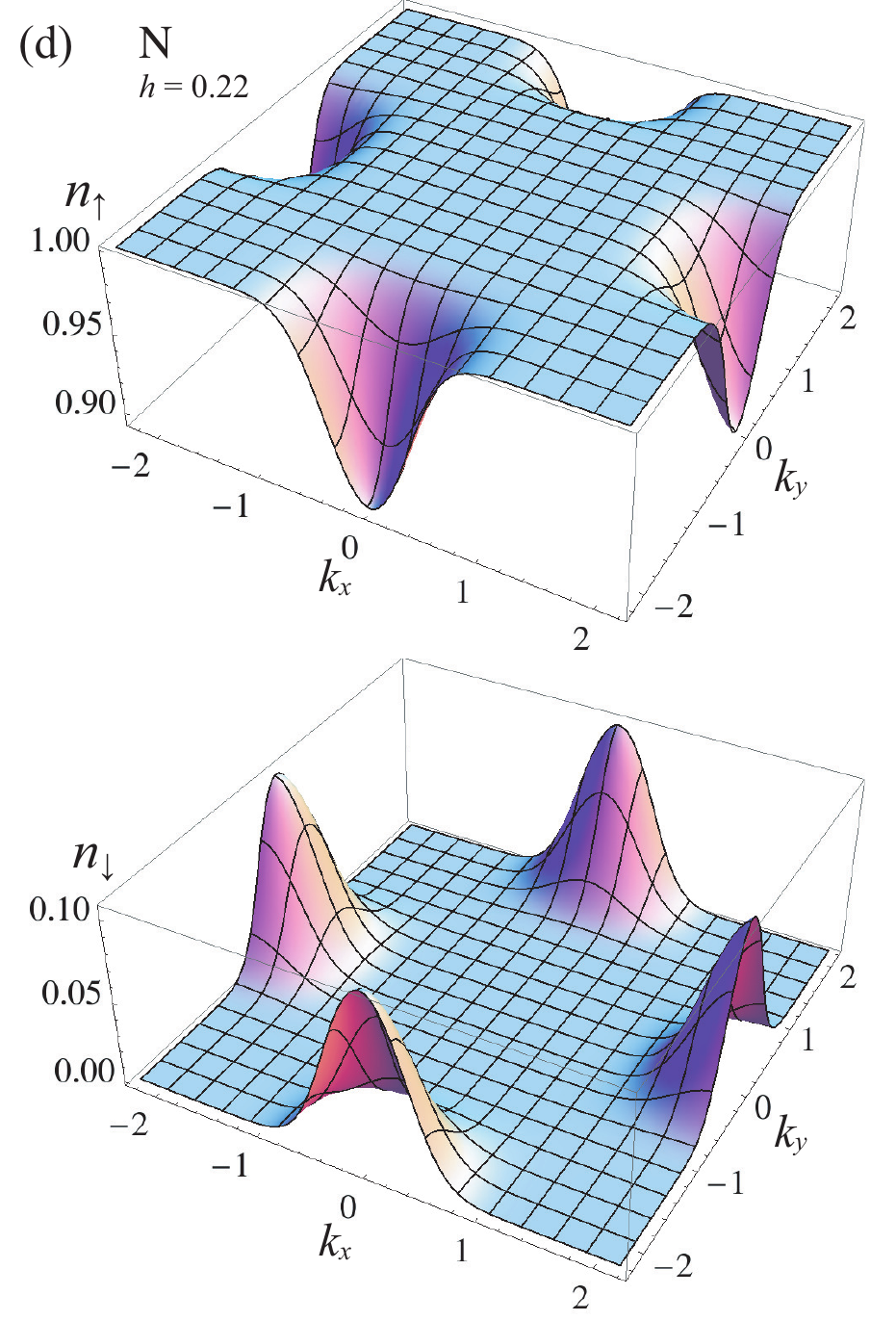}
\caption{Momentum distributions of the spin-up (top) and spin-down (bottom) particles averaged over the three bands as functions of $k_x$ and $k_y$ in a lattice with parameters $V_0=10$ and $\theta\approx0.556\pi$ for the points marked in Fig.\ref{phasediag_real}(a). The temperature is $T=0.01$, the interaction is $U=0.10$ and the chemical potential differences (a) $h=0.10$ (SF$_0$), (b) $h=0.14$ (SF$_\pi^1$), (c) $h=0.17$ (SF$_\pi^2$) and (d) $h=0.22$ (normal state).}\label{momentumU010T001}
\end{figure*}

\begin{figure*}
\includegraphics[width=0.24\textwidth]{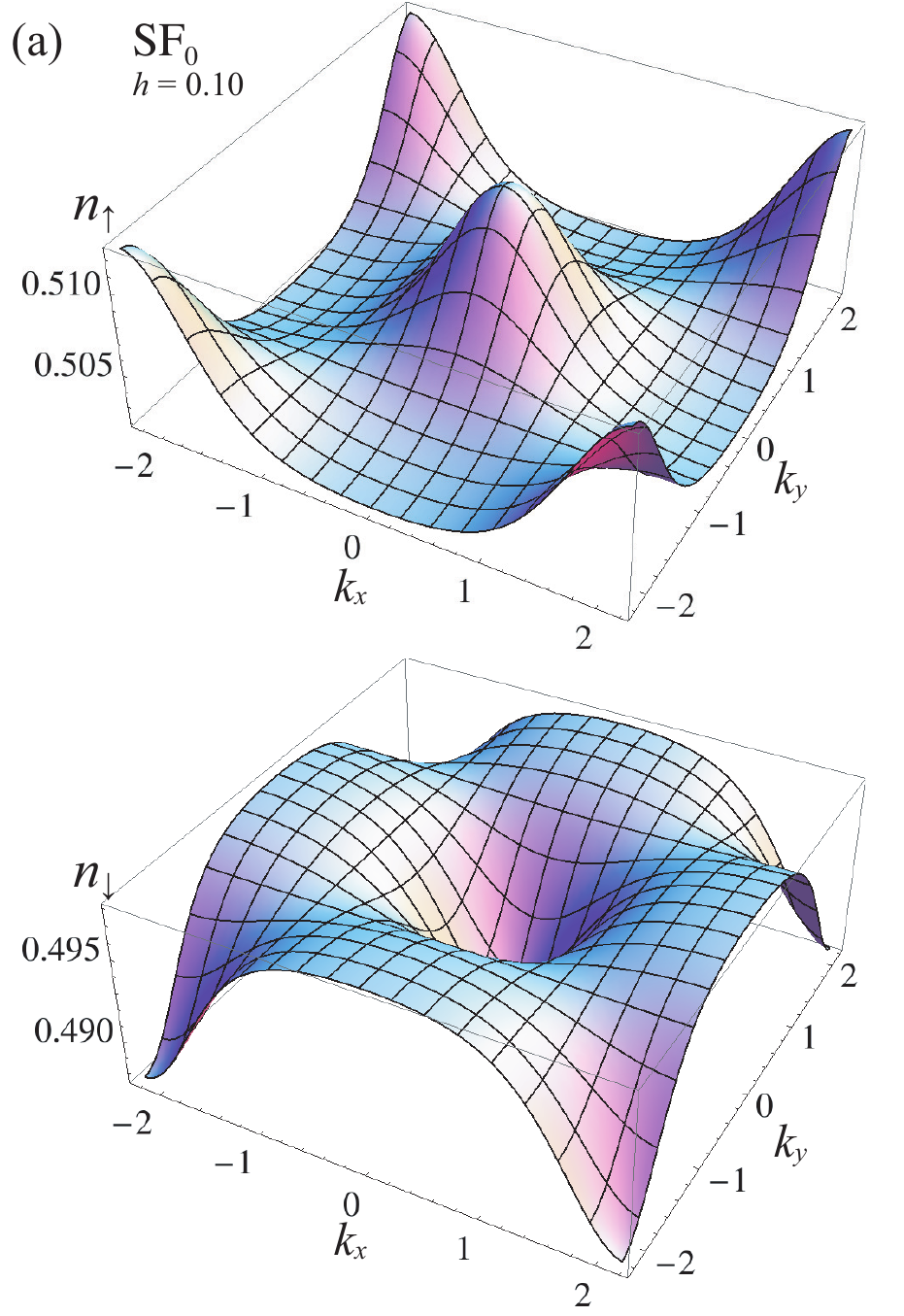}
\includegraphics[width=0.24\textwidth]{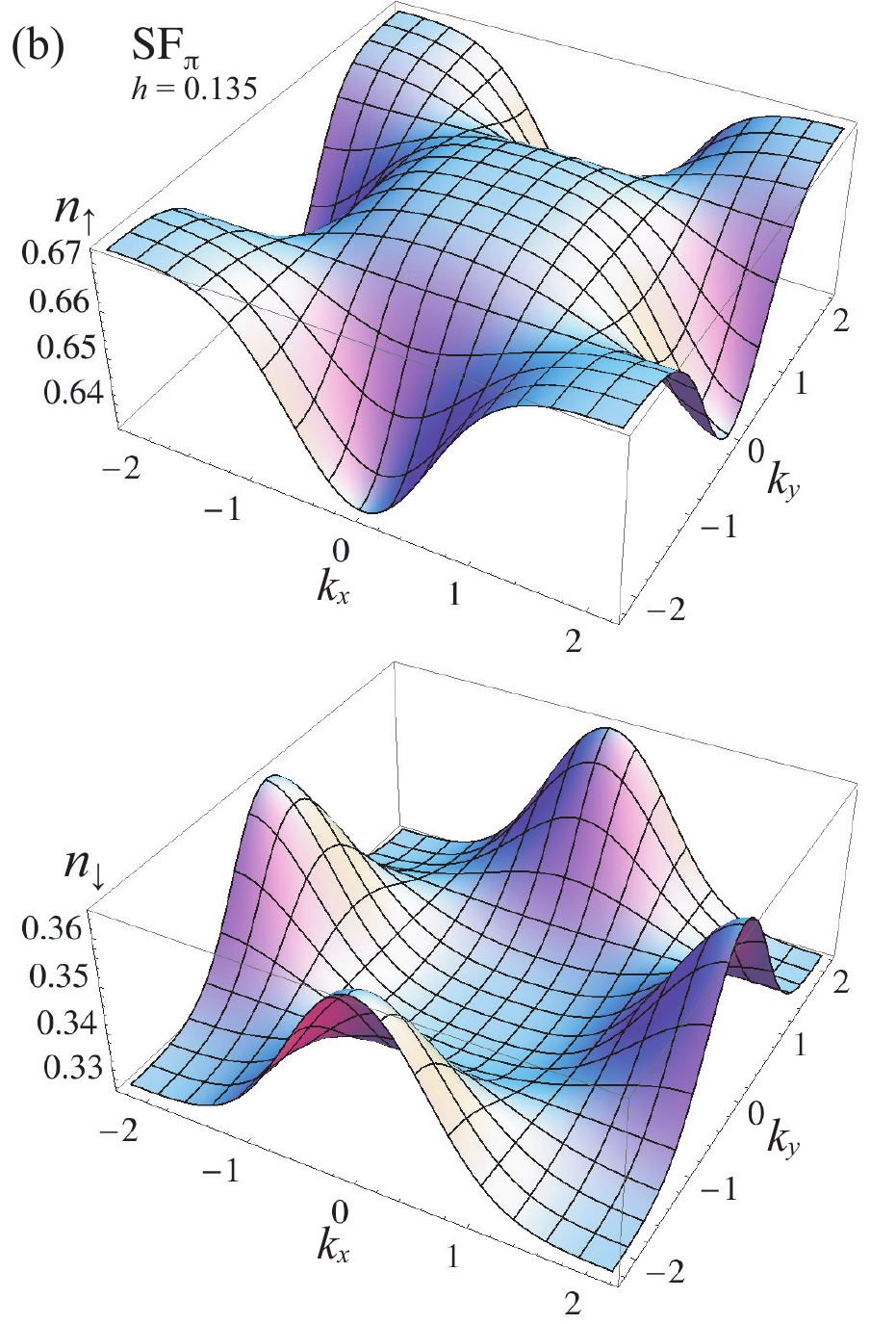}
\includegraphics[width=0.24\textwidth]{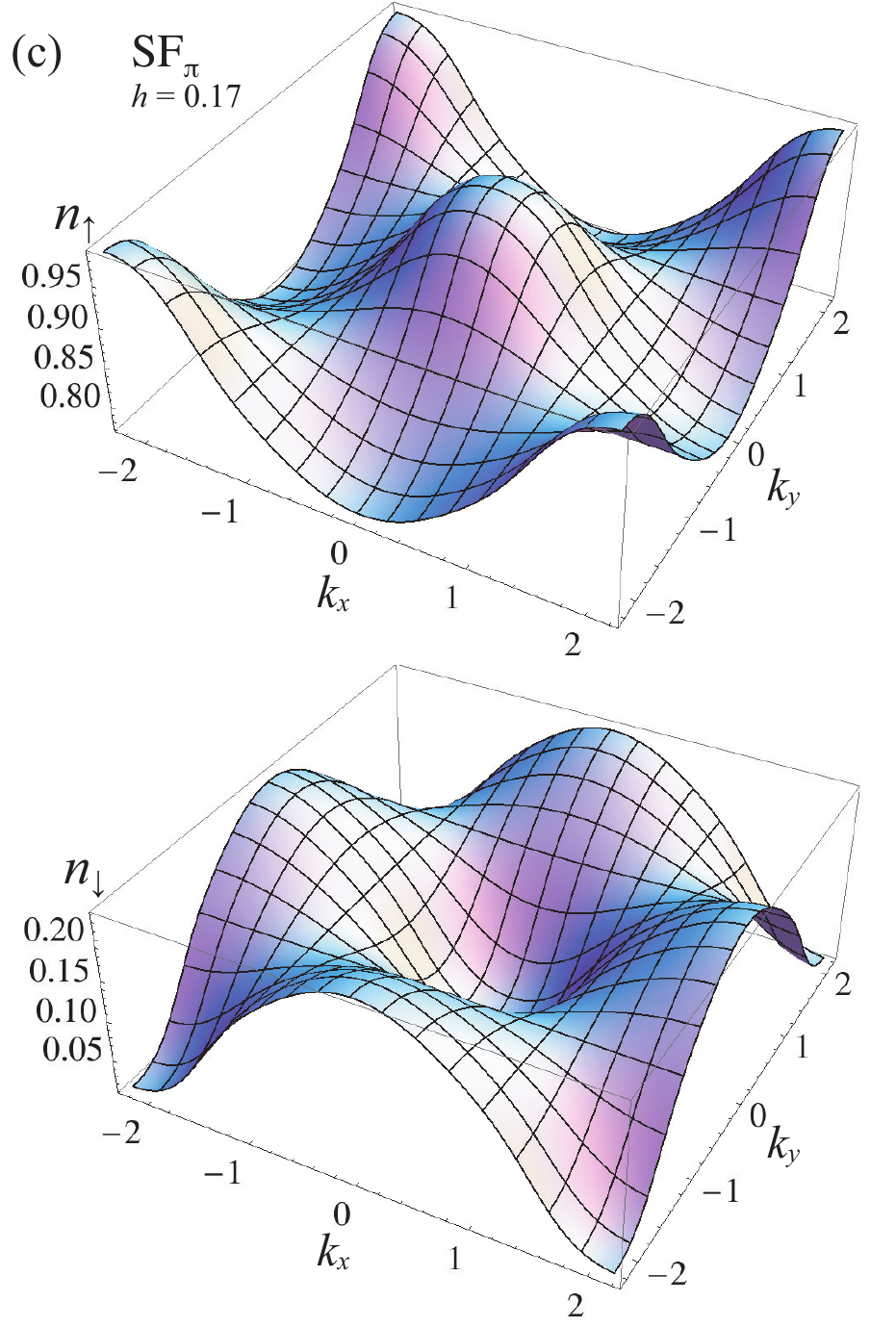}
\includegraphics[width=0.24\textwidth]{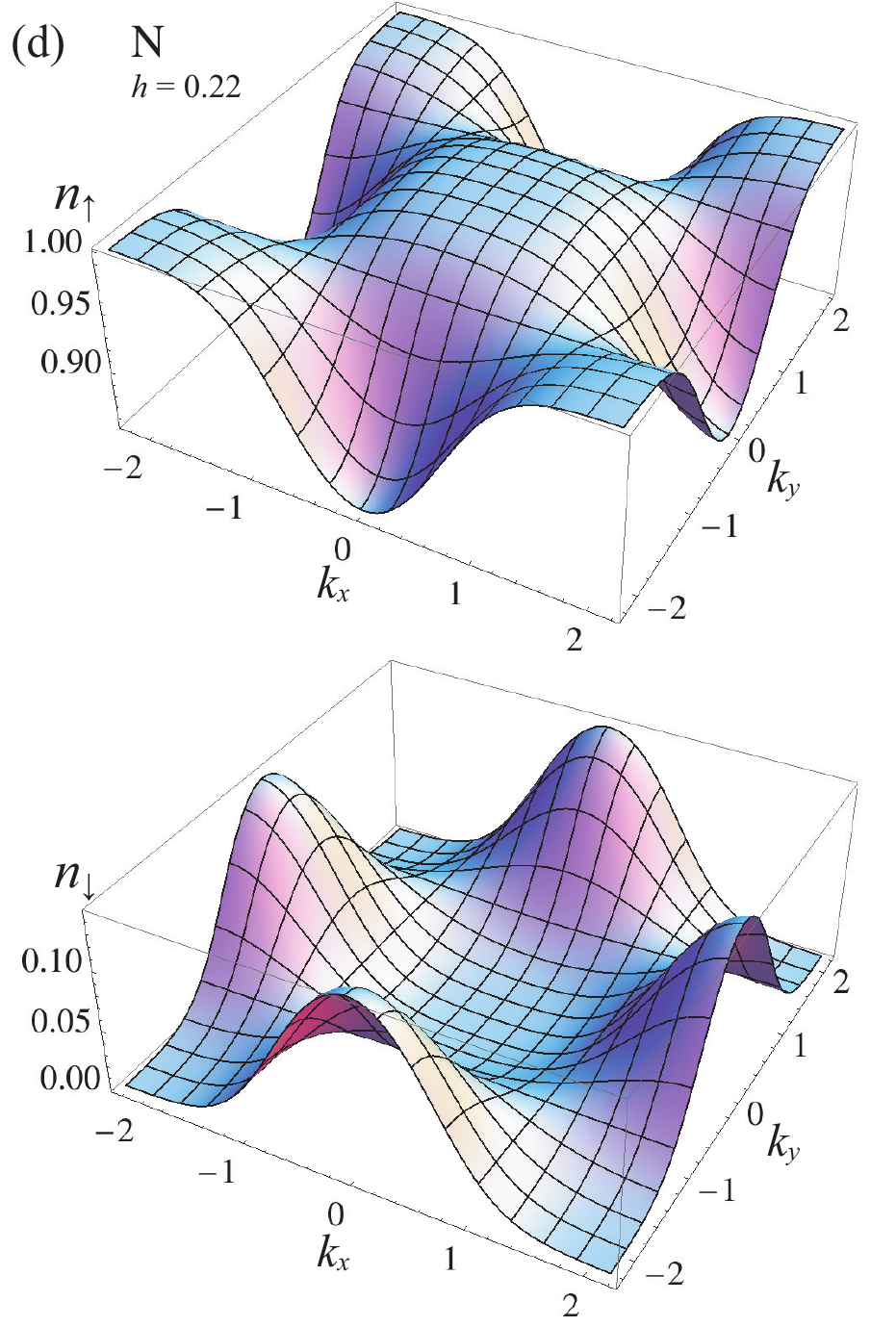}
\caption{The same as in Fig.~\ref{momentumU010T001} but at temperature $T=0.03$, corresponding to the points in Fig.\ref{phasediag_real}(b). Again the interaction is $U=0.10$ and now the chemical potential differences are (a) $h=0.10$ (SF$_0$), (b) $h=0.135$ (SF$_\pi^1$), (c) $h=0.17$ (SF$_\pi^2$) and (d) $h=0.22$ (normal state).}\label{momentumU010T003}
\end{figure*}

\subsection{Momentum distributions}
As a possible experimental signature of the SF$_0$ and SF$_\pi$ phases, we present the quasi-momentum distributions of the particles which can be observed experimentally. Since in such experiment, the original spin particles rather than the quasi-particles are observed, the particle distributions should be obtained by rotating the quasi-particle basis back to the original particle basis. This can be carried out by using the eigenvectors of $\mathbb{H}_{BCS}$ in Eq.~(\ref{Hamiltonianmatrix}), which form the transformation matrix $\mathbb{S}$ that diagonalizes $\mathbb{H}_{BCS}$. Then, the particle occupation number of the $i^\mathrm{th}$ state reads
\begin{equation}\label{momentum}
n_i(\mathbf{k})=\sum_j\frac{|\mathbb{S}_{ij}(\mathbf{k})|^2}{e^{\beta \omega_j(\mathbf{k})}+1},
\end{equation}
where $\omega_j(\mathbf{k})$ are the eigenvalues of $\mathbb{H}_{BCS}$ in Eq.\ref{Hamiltonianmatrix}. Within the six bands given in the above equation, there are only three independent distributions, since $n_{\mathbf{k}\uparrow}+n_{\mathbf{k}\downarrow}=1$ at half filling. Besides, since it is not possible to distinguish the original $s$ and $p$ bands of each particle in a measurement, we present the averaged occupations of the three bands for spin-up particles and for spin-down particles. 

Fig.\ref{momentumU010T001} shows the quasi-momentum distributions for the points marked in Fig.\ref{phasediag_real}(a), with interaction $U=0.10$, temperature $T=0.01$ and various values of the chemical potential difference $h$. The distributions show smooth changes as a function of the momenta. At zero temperature, these would be sudden jumps corresponding to the Fermi surfaces of the filled bands. It can be seen that the momentum distributions are very different for the different phases. Especially, the difference between the SF$_0$ and SF$_\pi$ phase is considerable.
Also the effect of the chemical potential difference on the densities can be seen quite clearly. At low $h$, in Fig.\ref{momentumU010T001}(a), the differences between the $\uparrow$ and $\downarrow$ distributions are very small, meaning that the densities are similar. In contrast, at a large value of $h$, in Fig.\ref{momentumU010T001}(d), the difference between the $\uparrow$ and $\downarrow$ distributions is very large, corresponding to a large polarization.
The various shapes of the distributions result from the interplay of the dispersions, being different for the various phases (Fig.\ref{dispersh}), the occupations of those levels, which depend on the chemical potential difference $h$, and the temperature.

The momentum distributions for the same interaction $U=0.10$, but at a higher temperature $T=0.03$ are shown in Fig.\ref{momentumU010T003} for various values of the chemical potential difference $h$, marked in Fig.\ref{phasediag_real}(b). It can be seen that the qualitative differences between the momentum distributions for the different phases are still there, although a bit smoothened out compared to the $T=0.01$ distributions. However, the variations in the momentum distributions are now much larger, making the experimental observation of these interesting phases possible. 

At even higher temperatures the variations in the distributions are even larger, but the qualitative differences between them for the various phases are then completely smoothened out. 
The quasi-momentum distributions also change with varying the interaction strength $U$. For the same phases the distributions are qualitatively the same as the ones depicted in Fig.\ref{momentumU010T001} and in Fig.\ref{momentumU010T003}. However, the variations in the distributions for both spin components become smaller at larger interaction strength and larger for smaller interactions $U$.

\section{Conclusion and Outlook}
In conclusion, we studied lattices populated by two-component fermions occupying both $s$ and $p$ orbital states in both one and two dimensions. We showed how the system in two dimensions can be mapped to a Lieb lattice. In 1D we used a simple mean-field calculation without including the full dispersions of the particles and determined the phase diagram, which shows two different superfluid phases. One superfluid phase is a homogeneous superfluid phase, while the other one is an inhomogeneous superfluid phase, so-called $\pi$ phase, having similarities with the LO superfluid phase. Consequently, we calculated the full thermodynamic potential for an experimantally realizable two-dimensional lattice within a mean-field theory and find a similarly rich phase diagram. 
Also, we calculated the momentum distributions for the two spin components in the system, which could be observed experimentally. 

Due to the hybridization of the $s$ and $p$ bands a flat band appears in the system. Flat bands can be related to many topological properties \cite{heikkila_dimensional_2011,volovik_flat_2011,Sebastiano_2015} and may be responsible for high $T_c$ surface superconductor \cite{kopnin_high-temperature_2011}. In future research, we will focus on the flat dispersion entering in this theory and the role of a flat band on pairing instabilities.

\begin{acknowledgments}
This work was supported by the Academy of Finland through its
Centres of Excellence Programme (Projects No. 263347,
No. 251748, No. 135000, and No. 272490) and by the
European Research Council (ERC-2013-AdG-340748-CODE).
\end{acknowledgments}

\bibliographystyle{apsrev}
\bibliography{./Multiband_literature}

\end{document}